\documentclass[prd,preprintnumbers,twocolumn,eqsecnum,floatfix,a4paper,nofootinbib,superscriptaddress]{revtex4-1}

\usepackage{color}
\usepackage{amsmath,amssymb,graphicx}
\usepackage{bm,lipsum}
\usepackage{times}
\usepackage{microtype}
\usepackage[utf8]{inputenc}
\usepackage{booktabs}
\usepackage{subfigure}
\usepackage[normalem]{ulem}
\usepackage[varg]{txfonts}
\usepackage{bm,graphics,graphicx,epsfig,color,ulem}
\usepackage[colorlinks, pdfborder={0 0 0}]{hyperref}
\definecolor{LinkColor}{rgb}{0.75, 0, 0}
\definecolor{CiteColor}{rgb}{0, 0.5, 0.5}
\definecolor{UrlColor}{rgb}{0, 0, 0.75}
\hypersetup{linkcolor=LinkColor}
\hypersetup{citecolor=CiteColor}
\hypersetup{urlcolor=UrlColor}
\allowdisplaybreaks
\def\linechange{\nonumber\\&&}
\newcommand{\msun}{$\text{M}_{\odot}$}

\begin{document}


\title{Testing the multipole structure and conservative dynamics of compact binaries using
gravitational wave observations: The spinning case}
\author{Shilpa Kastha}\email{shilpakastha@imsc.res.in}
\affiliation{The Institute of Mathematical Sciences, C. I. T. Campus, Taramani, Chennai, 600113, India}
\affiliation{Homi Bhabha National Institute, Training School Complex, Anushakti Nagar, Mumbai, 400094, India}

\author{Anuradha Gupta} \email{axg645@psu.edu}
\affiliation{Institute for Gravitation and the Cosmos, Department of Physics, Penn State University, University Park PA 16802, USA}

\author{K. G. Arun} \email{kgarun@cmi.ac.in} 
\affiliation{Chennai Mathematical Institute, Siruseri, 603103, India}
\affiliation{Institute for Gravitation and the Cosmos, Department of Physics, Penn State University, University Park PA 16802, USA}

\author{B. S. Sathyaprakash} \email{bss25@psu.edu} 
\affiliation{Institute for Gravitation and the Cosmos, Department of Physics, Penn State University, University Park PA 16802, USA}
\affiliation{Department of Astronomy and Astrophysics, Penn State University, University Park PA 16802, USA}
\affiliation{School of Physics and Astronomy, Cardiff University, Cardiff, CF24 3AA, United Kingdom}

\author{Chris Van Den Broeck}\email{vdbroeck@nikhef.nl}
\affiliation{Nikhef - National Institute for Subatomic Physics, Science Park 105, 1098 XG Amsterdam, The Netherlands}
\affiliation{Van Swinderen Institute for Particle Physics and Gravity, University of Groningen, Nijenborgh 4, 9747 AG Groningen, The Netherlands}
\date{\today}

\begin{abstract}
 In an earlier work [S. Kastha et al.,  PRD
{\bf 98}, 124033 (2018)], we developed the {\it parametrized multipolar
gravitational wave phasing formula} to test general relativity, for the
non-spinning compact binaries in quasi-circular orbit. In this paper, we
extend the method and include the important effect of spins in the
inspiral dynamics. Furthermore, we consider parametric scaling of
PN coefficients of the conserved energy for the compact binary,
resulting in the parametrized phasing formula for non-precessing
spinning compact binaries  in quasi-circular orbit. We also compute
the projected accuracies with which the second and third generation
ground-based gravitational wave detector networks as well as the planned
space-based detector LISA  will be able to measure the multipole
deformation parameters and the  binding energy parameters. Based on different source configurations, we find that a network of third-generation detectors would have comparable ability to that of LISA in constraining the conservative and dissipative dynamics of the compact binary systems. This parametrized multipolar waveform would be extremely useful not only in deriving the first upper limits on any deviations of the multipole and the binding energy coefficients from general relativity using the gravitational wave detections, but also for science case studies of next generation gravitational wave detectors.
	
\end{abstract}
\maketitle

\section{Introduction}

Mergers of compact binaries are unique probes of the predictions of general relativity (GR) in the
strong-gravity regime~\cite{WillLR05,SathyaSchutzLivRev09,YunesSiemens2013,GairLivRev,BertiCQGRev}.
The gravitational wave (GW) detections made so
far~\cite{Discovery,GW151226,GW170104,O1BBH,GW170608,GW170814,GWTC1} by advanced
LIGO~\cite{aLIGO} and advanced Virgo~\cite{aVirgo}, have been used in
various ways to test GR by employing different
methods~\cite{TOG,O1BBH,GW170104,YYP2016,GW-GRB170817,BNSTGR} to find
very good agreement with the predictions of GR within the statistical uncertainties. With
several more of such events expected to be detected in the future observing runs,
developing efficient methods to carry out such tests will play a central
role in extracting the best  science from these observations.
Ongoing developments  of the science case for third-generation ground-based detectors such as
Einstein Telescope~\cite{ET} and Cosmic Explorer~\cite{CEDwyer}, and space-based LISA
interferometer~\cite{LPF, LISA2017} further motivates developing generic methods to test GR
using GWs.

There are a wide variety of tests proposed in the literature to
assess GR using GW observations. These are often broadly classified as model independent tests (or
theory-agnostic tests) and theory-dependent tests. Parametrized tests of
GR \cite{AIQS06a,AIQS06b,Arun2012,YunesPretorius09,MAIS10,TIGER,
Li:2011cg,Meidam:2017dgf}, Parametrized post-Einsteinian
framework~\cite{YunesPretorius09,PPE2011} and inspiral-merger-ringdown
consistency tests~\cite{IMRConsistency} are examples of the first kind
whereas and the model dependent tests include tests aimed at looking for
signatures of a specific alternative theory (or a class of alternative
theories) such as those suggested in
Refs.~\cite{WillBD94,KKS95,Will98,MYW11}.

Recently, we proposed a new
theory-agnostic test  to probe the multipolar structure of compact binaries in
GR~\cite{Kastha2018}.  The basic idea is to ask using GW
observations how well we can infer the multipole structure of the gravitational field
of the compact binary and search for potential deviations. In order to answer this question, we computed a parametrized gravitational waveform
model  explicitly keeping track of the contributions to the
gravitational waveform from different radiative-multipole
moments of the
compact binary following the formalism developed in
Refs.~\cite{Th80,BDI95,BDIWW95,BIWW96,BIJ02,BDEI04}. This prescription
is built on the post-Newtonian (PN)  approximation developed for compact binary systems 
with non-spinning component
masses in quasi-circular orbits. By introducing seven independent
parameters associated to the deviation of the seven radiative-multipole
moments from GR, we re-derived the GW flux.
This parametrized multipolar waveform facilitates tests of GR in a model independent way with GW observations~\cite{Kastha2018}. We computed the projected accuracies on the measurements of these multipole coefficients for various ground-based and space-based detectors~\cite{Kastha2018}.

There is a strong astrophysical evidence that stellar mass black hole (BH)
binaries~\cite{Abramowicz:2001bi, Gou:2013dna} as well as super-massive
BH binaries \cite{Reynolds:2013rva} may have highly spinning binary
constituents. The spins of the compact binary components affect the binary
dynamics and give rise to a very different radiation profile as compared
to their non-spinning counterparts. Hence a physically realistic
waveform model should account for the spin dynamics of compact binaries.
Within the PN formalism, the gravitational waveform has been calculated
considering the point masses with arbitrary spins up to a very high
accuracy~\cite{KWWi93,Apos95,K95,Poisson97,Miketal05,BBuF06, ABFO08,
BBF2011,BMFB2012, BMB2013, BFMP2015, Bliving2014,Hartung2011_1, MBFB2012,Levi_2016,M3B2013, Marsat2014, BFH2012,WWi96, Hartung2011_2, Levi2010}.
Hence, in this paper, we extend our
parametrized multipolar GW energy flux as well as PN waveform model,
presented in Ref.~\cite{Kastha2018}, with spin-orbit and spin-spin contributions 
from binary components. We assume that the component spins are either aligned or anti-aligned with respect to the orbital angular momentum of the binary which is inspiraling in quasi-circular orbit. 
 Here, in addition to the multipolar structure, we present the
phasing formula which also parametrizes the conservative dynamics of the
binary. This is achieved by introducing free parameters at each PN order
in the binding energy expression which take value unity
in GR, by definition.

Having included the effects of spins in our parametrized test of
multipole structure, we use Fisher information matrix based parameter
estimation scheme to compute projected bounds on the various multipolar parameters.
Along with the complete study on the bounds of the multipolar parameters, 
we also provide the bounds on the parameters associated to conservative sector 
for the first time in this paper. 
We consider GW observation through networks of the various second 
and third generation ground-based detectors as well the proposed space-based 
LISA mission~\cite{LISA2017}. Inclusion of spin effects not only increases the 
dimensionality of the parameter space but also degrades the measurement accuracy of parameters.
We find that a network of third-generation ground-based detectors and
the space-based LISA mission would
have comparable sensitivity to  detect potential deviations in the
multipolar structure of compact binaries.

This paper is organized as follows. In Sec.~\ref{parameterized GW flux}
we discuss our computational scheme for the multipolar parametrized
gravitational wave energy flux. In Sec.~\ref{waveform} we explore the
modifications in the parametrized frequency domain (TaylorF2) waveform
due to the various contributions from spins. Thereafter, in
Sec.~\ref{sec:pe} we briefly describe the parameter estimation
techniques we use in this paper.  Section~\ref{detector config} provides
a detailed description about the various GW detector configurations used for our analysis. In Sec.~\ref{sec:results} we discuss
the bounds on the multipole coefficients for various GW
detectors and  Sec.~\ref{sec:conc} presents our concluding remarks.

\section{Parametrized gravitational wave energy flux}\label{parameterized GW flux}

During the inspiral phase of the compact binary dynamics, the radiation reaction time scale is much longer than the time scale for orbital motion. Due to this separation of time scales,  two vital ingredients for computing the phase evolution are the {\it conserved orbital energy} of the binary and the {\it gravitational wave energy flux} from the system. While the former characterizes the conservative dynamics of the binary, the latter describes the dissipative dynamics.

 The computation of the multipolar parametrized flux formula makes use of
the entire machinery of the Multipolar post-Minkowskian and
post-Newtonian formalism developed over past several
years~\cite{BD86,BD87,BS89,BDI95,B95,BIJ02,DJSequiv,BDEI04,BBuF06} (see
\cite{Bliving} for a review.)
Following Ref.~\cite{Kastha2018}, we use the GW energy flux parametrized
in terms of the various radiative multipole moments of compact binary
while including contributions from the spins of the binary components in
quasi-circular orbits. More explicitly, to capture the generic
deviations from GR, parametric deviations are introduced at the level of
mass-type ($U_L$) and current-type ($V_L$) radiative multipole moments through simple scaling relationships of the
kind
\begin{align}
\label{scaled-Mmoments}
U_L\rightarrow \mu_l\,U_L^{\rm GR},\\
V_L \rightarrow \epsilon_l \, V_L^{\rm GR}\label{scaled-Cmoments},
\end{align}
where $\mu_l=1+\delta U_L/U_L^{\rm GR}$ and  $\epsilon_l=1+\delta
V_L/V_L^{\rm GR}$ take the value unity in GR. In this paper we focus on
the contributions to the flux from spin angular momentum of the binary
components and hence quote only the spin-dependent part of the
parametrized GW energy flux  which may be added to the non-spinning
results of \cite{Kastha2018} to get the complete phasing. Among the few different approaches to consider the PN spin corrections to the conservative dynamics as well as gravitational radiation from a compact binary system, we adopt the PN iteration scheme in harmonic coordinates~\cite{Bliving2014} to obtain spin contributions to the radiative moments in GR which we further rescale as described in Eqs.~(\ref{scaled-Mmoments})-(\ref{scaled-Cmoments}).

We closely follow the prescription given in Refs.~\cite{BBuF06,
BBF2011, BMB2013, BFMP2015, Bliving2014} to account for the
contributions to the conservative and dissipative sectors of the compact
binary dynamics from the individual spins. In our notation, the
individual spins of the component masses, $m_1$ and $m_2$ are
$\mathbf{S_1}$ and $\mathbf{S_2}$ with quadrupolar polarisabilities
$\kappa_1$ and $\kappa_2$, respectively, which are unity for
Kerr black holes. We denote the total mass for the system to be $m=m_1+m_2$, relative mass difference, $\delta=(m_1-m_2)/m$ and the symmetric mass ratio, $\nu=m_1 m_2/m^2$. Furthermore following the usual notation, we present our results in terms of the symmetric combination of the quadrupolar polarisabilities, $\kappa_+=\kappa_1+\kappa_2$ and the anti-symmetric combination, $\kappa_-=\kappa_1-\kappa_2$. Our results are expressed in the Center of Mass frame where the spin variables $\mathbf{S}$ and $\mathbf{\Sigma}$ have the following relations with the spins of each of the constituent masses of the binary,

\begin{eqnarray}
\mathbf{S}&&=\mathbf{S_1}+\mathbf{S_2}\,,\\
\mathbf{\Sigma}&&=m\Bigg(\frac{\mathbf{\Sigma_2}}{m_2}-\frac{\mathbf{\Sigma_1}}{m_1}\Bigg)\,,
\end{eqnarray}
and $S_L=\mathbf{S\cdot L}$ and $\Sigma_L=\mathbf{\Sigma\cdot L}$ are the projections along the direction of orbital angular momentum.

Depending on the order of spin corrections, the GW flux schematically has the following structure, 
\begin{equation}
\mathcal{F}= \mathcal{F}_{\rm NS} + \mathcal{F}_{\rm SO} + \mathcal{F}_{\rm SS}+ \mathcal{F}_{\rm SSS}+....\label{flux_spin},
\end{equation}
where $\mathcal{F}_{\rm NS}$ is the non-spinning contribution computed
in Eq. (2.8) of Ref.~\cite{Kastha2018}, $\mathcal{F}_{\rm SO} $ is the
spin-orbit (SO) contribution which linearly depends on the spins, and
$\mathcal{F}_{\rm SS}$ is quadratic in spins arising due to the
spin-spin (SS) interactions. Similarly $\mathcal{F}_{\rm SSS}$ denotes
the cubic-in-spin effects on the GW energy flux. Here we report the
parametrized multipolar flux accounting for spin-orbit effects up to 3.5PN
order and quandratic-in-spin contributions up to 3PN order. 
We do not provide the  cubic spin  and the partial
quadratic-in-spin contribution at 3.5PN order. The non-spinning flux
computed in Ref.~\cite{Kastha2018} should be added to these to obtain
the total flux. We provide explicit expressions for the
spin-orbit and quadratic-in-spin contributions to multipolar parametrized GW fluxes 
in the following subsections.

\subsection{Spin-orbit contribution}

Considering the leading order spin corrections to the multipole moments
as well as in the equation of motion (EOM) and following the same
technique as prescribed in Refs.~\cite{BBuF06, BBF2011, BMB2013}, we
re-compute the parametrized SO part of the energy flux, which is given
as

\begin{widetext}
 \begin{eqnarray}
 \mathcal{F}_{\rm SO}=&&\frac{32}{5}\frac{c^5 }{G} \nu ^2 \mu_2^2 x^5 \Bigg\{
 	\frac{x^{3/2}}{G m^2}\Bigg(- 4 S_L+\delta\Sigma_L\Big[-\frac{4}{3}+\frac{\hat{\epsilon}_2^2}{12}\Big]\Bigg)
 	+	\frac{x^{5/2}}{G m^2}\Bigg(S_L\Big[ \frac{316}{63}-\frac{514}{63}\nu-\hat{\mu}_3^2\Big(\frac{598}{63}-\frac{2392 }{63}\nu\Big)-\hat{\epsilon}_2^2\Big(\frac{43}{126}-\frac{86 }{63}\nu\Big)\linechange
 	+\hat{\epsilon}_3^2\Big(\frac{20}{63}-\frac{20}{21}\nu\Big)\Big]
 	+\delta\Sigma_L\Big[
 	\frac{208}{63}-\frac{10}{9}\nu-\hat{\mu}_3^2\Big(\frac{1025}{252}-\frac{1025 }{84}\nu\Big)-\hat{\epsilon}_2^2\Big(\frac{367}{1008}-\frac{11 }{18}\nu\Big)	 		
 	+\hat{\epsilon}_3^2\Big(\frac{20}{63}-\frac{20}{21}\nu\Big)
 	\Big]\Bigg) \linechange
 	+	\frac{\pi x^{3}}{G m^2}\Bigg(-16S_L+\delta\Sigma_L\Big[-\frac{16}{3}+\frac{\hat{\epsilon}_2^2}{6}\Big]\Bigg) 
 	+	\frac{x^{7/2}}{G m^2}\Bigg(S_L\Big[\frac{58468}{1323}+ \frac{154424}{1323}\nu +\frac{3494}{1323} \nu ^2+ \hat{\mu}_3^2\Big(\frac{120121 \nu ^2}{1134}-\frac{345665 \nu }{1512}+\frac{65491}{1296}\Big)\linechange
 	+ \hat{\mu}_4^2\Big(-\frac{272392 \nu ^2}{1323}+\frac{544784 \nu }{3969}-\frac{272392}{11907}\Big)+ \hat{\epsilon}_2^2\Big(-\frac{1534 \nu ^2}{3969}-\frac{1165 \nu }{2646}+\frac{2131}{15876}\Big)+ \hat{\epsilon}_3^2\Big(-\frac{7300 \nu ^2}{567}+\frac{7150 \nu }{567}-\frac{1556}{567}\Big)\linechange
 	+ \hat{\epsilon}_4^2\Big(\frac{5741 \nu ^2}{882}-\frac{5741 \nu }{1176}+\frac{5741}{7056}\Big)
 	\Big]
 	+\delta\Sigma_L\Big[\frac{28423 \nu ^2}{3969}+\frac{366697 \nu }{7938}+\frac{49844}{3969} + \hat{\mu}_3^2\Big(\frac{319661 \nu ^2}{18144}-\frac{811795 \nu }{9072}+\frac{253385}{9072}\Big) \linechange
 	+ \hat{\mu}_4^2\Big(-\frac{3184 \nu ^2}{49}+\frac{7960 \nu }{147}-\frac{1592}{147}\Big) 
 	+ \hat{\epsilon}_2^2\Big(-\frac{41471 \nu ^2}{127008}-\frac{37585 \nu }{31752}+\frac{14383}{63504}\Big)+ \hat{\epsilon}_3^2\Big(-\frac{490 \nu ^2}{81}+\frac{5140 \nu }{567}-\frac{188}{81}\Big) \linechange
 	+ \hat{\epsilon}_4^2\Big(\frac{5741}{7056}-\frac{28705 }{7056}\nu+\frac{5741 }{1176}\nu^2\Big)
 	\Big]\Bigg) 	
 	\Bigg\}\,.\label{flux_SO}
\end{eqnarray}
\end{widetext}

Spin-orbit corrections to the flux first appear at 1.5PN order due to
spin-dependent terms in mass quadrupole moments at 1.5PN order and current
quadrupole moment at 0.5PN order. Hence the leading order SO corrections bring in the  $\mu_2$ and $\epsilon_2$  in the parametrized GW flux at 1.5PN. As clearly stated in Ref.~\cite{BBuF06}, at 2.5PN order the SO contributions come from mass- and current-type quadrupole and octupole moments, which is evident from Eq.~(\ref{flux_SO}) since only $\mu_2,\mu_3,\epsilon_2$ and $\epsilon_3$ are present up to 2.5PN order. At 3PN order, the spin dependences come from the 1.5PN tail integral performed on mass quadrupole moment and the 2.5PN tail integral performed on current quadrupole moment~\cite{BBF2011}. Hence at 3PN order only $\mu_2$ and $\epsilon_2$ are present. As we go to higher order we find that at 3.5PN order, $\mu_4$ and $\epsilon_4$ are also present along with the lower order coefficients.
As a check on the calculation, in the limit $\mu_2=\mu_3=\mu_4=\mu_5=\epsilon_2= \epsilon_3=\epsilon_4=1$, Eq.~(\ref{flux_SO}) reduces to Eq.~(4) of
Ref.~\cite{BBuF06}.

\subsection{Spin-spin contribution}

Quadratic-in-spin corrections first appear at 2PN order to the GW flux and the waveform (see Refs.~\cite{KWWi93,Miketal05,Racine2008,K95,WWi96} for details), whereas SS terms at 3PN are first calculated in Ref.~\cite{BFMP2015}. 

Along with the terms quadratic-in-spin in the EOM, the complete SS contributions to the flux are generated from the four leading multipole moments, $I_{ij},I_{ijk},J_{ij}$ and $J_{ijk}$. Hence $\mathcal{F}_{\rm SS}$ is completely parametrized by $\mu_2, \mu_3, \epsilon_2$ and $\epsilon_3$ (see Eq.~(\ref{flux_SS})). We have also written the SS contribution at 3.5PN order arising from the two leading order tail integrals performed on mass and current quadrupole moments. However, at 3.5PN order SS contributions are partial. Hence these contributions will be neglected for the waveform computations.

\begin{widetext}
\begin{eqnarray}
 \mathcal{F}_{\rm SS}=&&\frac{32}{5}\frac{c^5 }{G} \nu ^2 \mu_2^2 x^5  \frac{1}{G^2 m^4}\Bigg\{
 	x^2\Bigg(S_L^2\Big[4+2\kappa_+\Big]+S_L \Sigma_L\Big[2\kappa_+\delta+4\delta-2\kappa_{-}\Big]+\Sigma_L^2\Big[\frac{\hat{\epsilon}_2^2}{16}+\kappa_+-\delta\kappa_--(4+2\kappa_+)\nu\Big]\Bigg)
 	\linechange
 	+x^3\Bigg(S_L^2\Bigg[-\frac{1198}{63}-\frac{46 \kappa_+}{7}+\frac{55 \delta  \kappa_-}{21}+\hat{\mu}_3^2\Big(\frac{1367}{168}+\frac{1367 \kappa_+}{336}-\frac{\delta\kappa_-}{1008}\Big)+\hat{\epsilon}_2^2\Big(\frac{1}{6} +\frac{\kappa_+}{12} - \frac{\delta \kappa_-}{18}\Big)+\frac{20}{63}\hat{\epsilon}_3^2 
 	\linechange
 	+\nu\Big(\frac{82}{7}+\frac{41 \kappa_+}{7}-\hat{\mu}_3^2\Big[\frac{1367}{42}+ \frac{1367 \kappa_+}{84}\Big]-\hat{\epsilon}_2^2\Big[\frac{2}{3} +\frac{\kappa_+}{3} \Big]\Big)\Bigg]
 	+ S_L \Sigma_L\Bigg[-\frac{193 \delta  \kappa_+}{21}-\frac{1436 \delta }{63}+\frac{193 \kappa_-}{21}\linechange
 	+\hat{\mu}_3^2\Big(\frac{293 \delta  \kappa_+}{72}+\frac{1367 \delta }{168}-\frac{293\kappa_-}{72}\Big)+\hat{\epsilon}_2^2\Big(\frac{5 \delta  \kappa_+}{36}-\frac{143 \delta }{252}-\frac{5 \kappa_-}{36}\Big) +\frac{40}{63}\delta\hat{\epsilon}_3^2 +\nu\Big(\frac{41 \delta  \kappa_+}{7}+ \frac{82 \delta }{7}-\frac{49 \kappa_-}{3}\linechange
 	+\hat{\mu}_3^2\Big[\frac{293  \kappa_-}{18}-\frac{1367 \delta }{42}-\frac{1367 \kappa_+ \delta}{84}\Big]-\hat{\epsilon}_2^2\Big[\frac{\delta  \kappa_+}{3}+\frac{2 \delta }{3}-\frac{5 \kappa_-}{9}\Big]\Big)\Bigg]
 	+\Sigma_L^2\Bigg[-\frac{26}{9}-\frac{193 \kappa_+}{42}+\frac{193 \delta  \kappa_-}{42}+\hat{\mu}_3^2\Big(\frac{293 \kappa_+}{144}-\frac{293 \delta  \kappa_-}{144}\Big)\linechange 
 	- \hat{\epsilon}_2^2\Big[\frac{31}{56}-\frac{5 \kappa_+}{72}+\frac{5 \delta  \kappa_-}{72}\Big]+\frac{20}{63}\hat{\epsilon}_3^2 +\nu\Big(\frac{1562}{63}+\frac{619 \kappa_+}{42} -\frac{233 \delta  \kappa_-}{42} -\hat{\mu}_3^2\Big[\frac{1367}{168}+\frac{12305 \kappa_+}{1008}-\frac{8203 \delta  \kappa_-}{1008}\Big]\linechange 
 	+\hat{\epsilon}_2^2\Big(\frac{167}{168}-\frac{13 \kappa_+}{36}+\frac{2 \delta  \kappa_-}{9}\Big)-\frac{80}{63}\hat{\epsilon}_3^2\Big)+\nu^2\Big(-\frac{41 \kappa_+}{7}-\frac{82}{7} +\hat{\mu}_3^2\Big[\frac{1367}{42}+ \frac{1367 \kappa_+}{84}\Big]+\hat{\epsilon}_2^2\Big[\frac{2}{3}+\frac{\kappa_+}{3}\Big]
 	\Big)\Bigg]\Bigg)
 	\linechange
 	+\pi x^{7/2}\Bigg(S_L^2\Big[16+8\kappa_+\Big]+S_L \Sigma_L\Big[8\kappa_+\delta+16\delta-8\kappa_-\Big]+\Sigma_L^2\Big[ \frac{\hat{\epsilon}_2^2}{8}+4\kappa_+-4\delta\kappa_--(16+8\kappa_+)\nu\Big]\Bigg)\Bigg\}\,.\label{flux_SS}
\end{eqnarray}
\end{widetext}

As an algebraic check, in the limit, $\mu_2=\mu_3=\mu_4=\mu_5=\epsilon_2 =\epsilon_3=\epsilon_4=1$ for Eq.~(\ref{flux_SS}), we confirm the recovery of the accurate
expression for SS contribution to GW flux in GR reported in Eq.~(4.14) of Ref.~\cite{BFMP2015}.

\section{Parametrized multipolar gravitational wave phasing}\label{waveform}

The GW phase and its frequency evolution are obtained by using the energy conservation law which essentially balances the rate of change of conserved orbital energy $E$ and the emitted GW flux,

\begin{equation}
 \mathcal{F}=-\frac{d}{dt}\,E.
\end{equation} 
Hence an accurate model for conserved orbital energy is needed to obtain the GW phasing formula. 
 
In GR, for a non-spinning compact binary
inspiraling in quasi-circular orbit, the expression for the conserved
energy per unit mass is given in
Refs.~\cite{DJSdim,BDE04,DJSequiv,ABF01,BI03CM,itoh1,BFIJ02}, whereas
the SO corrections upto 3.5PN order and the SS corrections upto 3PN
order are quoted in Refs.~\cite{BBuF06, BBF2011, BMB2013, BFMP2015}. 

In alternative theories of gravity, along with the deformations at the level of multipole moments, we expect the conserved orbital energy to be deformed as well. In order to incorporate theses effects, we introduce free parameters $\alpha_k$, characterizing the deviations at different PN orders in the expression of conserved energy defined in GR for compact binaries in aligned (or anti-aligned)-spin configuration. For spin corrections to conservative dynamics we consider SO contributions upto 3.5PN order and SS contributions at 3PN order to the energy.
The 3.5PN closed-form expression for the parametrized conserved energy
reads as

\begin{widetext}\begin{eqnarray}\label{eq:Econs}
	E(v)&=&-\frac{1}{2} \nu \alpha_0 v^2   \Bigg[ 1 - 
	\Bigg (\frac{3}{4} + \frac{1}{12} \nu \Bigg ) \hat{\alpha_{2}}v^2 + \Bigg\{\frac{14}{3}S_L + 2\delta\Sigma_L\Bigg\}\frac{\hat{\alpha_{3}}}{G m^2}v^3 - \Bigg\{ \frac{27}{8} -\frac{19}{8} \nu  + \frac{1}{24}\nu^2 + \frac{S_L^2}{G^2m^4}(\kappa_+ +2) +\frac{S_L \Sigma_L}{G^2m^4}(\delta\kappa_+ +2\delta - \kappa_-) \nonumber \\ 
	&&+\frac{\Sigma_L^2}{G^2m^4}\Big(\frac{1}{2}\kappa_+  - \frac{\delta}{2}\kappa_- - \nu [\kappa_+ + 2]\Big) \Bigg\} \hat{\alpha_{4}}v^4 +\Bigg\{\Big[11-\frac{61}{9}\nu\Big]S_L + \Big[3-\frac{10}{3}\nu\Big]\delta\Sigma_L\Bigg\}\frac{\hat{\alpha_{5}}}{Gm^2} v^5- \Bigg\{ \frac{675}{64}
	- \Big(\frac{34445}{576}-\frac{205 }{96}\pi^2\Big)\nu 
	+\frac{155 }{96}\nu^2\nonumber\\
	&&+\frac{35 }{5184}\nu^3
	+\frac{S_L^2}{G^2m^4}\Bigg(\Big[\frac{5}{3}\delta\kappa_- +\frac{25}{6}\kappa_+ -\frac{50}{9}\Big]-\nu\Big[\frac{5}{6}\kappa_+ +\frac{5}{3}\Big]\Bigg) +\frac{S_L\Sigma_L}{G^2m^4}\Bigg( \Big[\frac{5}{2}\delta\kappa_+ - \frac{25}{3}\delta -\frac{5}{2}\kappa_-\Big]-\nu\Big[\frac{5}{6}\delta\kappa_++\frac{5}{3}\delta +\frac{35}{6}\kappa_-\Big] \Bigg) \nonumber \\ 
	&&+\frac{\Sigma_L^2}{G^2m^4}\Bigg(\Big[\frac{5}{4}\kappa_+ -\frac{5}{4}\delta\kappa_- -5 \Big] -\nu\Big[\frac{5}{4}\kappa_+ +\frac{5}{4} \delta\kappa_- -10\Big] +\nu^2\Big[\frac{5}{6}\kappa_+ +\frac{5}{3}\Big]\Bigg)  \Bigg\} \hat{\alpha_6} v^6 + \Bigg\{\Big(\frac{135}{4}- \frac{367}{4}\nu+ \frac{29}{12}\nu^2 \Big)S_L \nonumber\\
	&& + \Big(\frac{27}{4}-39\nu+\frac{5}{4}\nu^2\Big)\delta\Sigma_L\Bigg\} \frac{\hat{\alpha_{7}}}{G m^2}v^7\Bigg ]\,,
	\label{eq:energy}
	\end{eqnarray}\end{widetext}
with $\hat{\alpha_i}=\alpha_i/\alpha_0.$
%
%
To obtain the gravitational waveform in frequency domain under the
stationary phase approximation~\cite{SatDhu91},  we use the
standard prescription outlined in Refs.~\cite{DIS00,BIOPS09}. The
important difference here is that we use the parametrized expressions
for the GW flux and  conserved energy given by Eq.~(\ref{flux_spin}) and (\ref{eq:energy}) respectively. 
Further we consider the amplitude to be at the leading quadrupolar order.  
The  standard {\it restricted} PN waveform in frequency domain, thus, reads as 
\begin{equation}
\tilde{h}_S(f)=\mathcal{A}\,\mu_2\,f^{-7/6}e^{i\psi_S(f)}\label{mod-TaylorF2},
\end{equation}
with $\mathcal{A}=\mathcal{M}_c^{5/6}/\sqrt{30}\pi^{2/3}D_L$; $\mathcal{M}_c=(m_1m_2)^{3/5}/(m_1+m_2)^{1/5}$ 
and $D_L$ are the chirp mass and luminosity distance. In the case of
LISA, to account for its triangular geometry, we multiply the
gravitational waveform by a factor of $\sqrt{3}/2$  while calculating
the Fisher matrix for LISA~\cite{Cutler98}. The parametrized multipolar phasing, $\psi_S(f)$, has the same structure as that of the energy flux (see Eq.~(\ref{flux_spin})). Schematically the parametrized phasing formula can be written as,

\begin{eqnarray}
	\psi_S(f)&=& 2\pi f t_c-\phi_c-\frac{\pi}{4}+\frac{3\alpha_0}{128 \nu v^5\mu_2^2} \Big[
     \psi_{\rm NS}(f)+\psi_{\rm SO}(f)\linechange 
     +\psi_{\rm SS}(f)
\Big],
\end{eqnarray}
where the parametrized non-spinning part, $\psi_{\rm NS}(f)$ is given by
Eq.~(A.2) in Ref.~\cite{Kastha2018}. Here we show only the SO and SS
parts: $\psi_{\rm SO}(f)$ and $\psi_{\rm SS}(f)$. As mentioned earlier,
we do not account for the partial contribution due to the spin-spin
interactions to the phasing formula at the 3.5PN order.

To evaluate the parametrized TaylorF2 phasing for aligned spin binaries, we use the conventional notations for the spin variables $(\bm{\chi_1}, \bm{\chi_2})$,  with the following re-definitions,


\begin{eqnarray}
\bm{\chi_1}=G m_1^2 \mathbf{S}_1,\\
\bm{\chi_2}=G m_2^2 \mathbf{S}_2.
\end{eqnarray}
Furthermore, we use $\chi_s=(\chi_1+\chi_2)/2$ and $\chi_a=(\chi_1-\chi_2)/2$ to present the phasing formula, where $\chi_1$ and $\chi_2$ are the projections of $\bm{\chi_1}$ and $\bm{\chi_2}$ along the orbital angular momentum, respectively. These spin variables have the following relations,
\begin{eqnarray}
&S_L&=G m^2[\delta\chi_a+(1-2\nu)\chi_s] \,,\\
&\Sigma_L&=-G m^2[\delta\chi_s+\chi_a] \,.
\end{eqnarray}

Finally we write down the expressions for $\psi_{\rm SO}$ and $\psi_{\rm SS}$, the main results of this paper, below

\begin{widetext}
	\begin{eqnarray}
	\psi_{\rm SO}=&& v^3\Bigg\{\Bigg[\frac{32}{3}+\frac{80}{3}\hat{\alpha_{3}}+\frac{1}{3}\hat{\epsilon}_2^2-\Bigg(\frac{32}{3}+\frac{40}{3}\hat{\alpha_{3}}+\frac{4}{3}\hat{\epsilon}_2^2\Bigg)\nu\Bigg]\chi_s+\Bigg[\frac{32}{3}+\frac{80}{3}\hat{\alpha_{3}}+\frac{1}{3}\hat{\epsilon}_2^2\Bigg]\delta\chi_a \Bigg\} 
	+ v^5\Bigg(1+3\log[v/v_{\rm LSO}]\Bigg)\Bigg\{\Bigg[- \frac{64160}{567}
	\linechange
	+\frac{93920}{567}\nu
	-\frac{1760}{189}\nu^2+\hat{\alpha_2}\Bigg(\frac{160}{9}-\frac{1280}{81}\nu-\frac{160}{81}\nu^2\Bigg)+\hat{\alpha_{3}}\Bigg(-\frac{85600}{567} +\frac{12400}{81}\nu-\frac{22000}{567}\nu^2\Bigg) +\hat{\alpha_{5}}\Bigg(-\frac{1120}{9} +\frac{16940}{81}\nu
	\linechange
	-\frac{280}{81}\nu^2\Bigg)
	+\Bigg(\frac{13670}{1701}- \frac{58090}{1701}\nu+\frac{13640}{1701}\nu^2+\hat{\alpha_{3}} \Bigg[\frac{68350}{1701}-\frac{34175}{189}\nu+\frac{136700}{1701}\nu^2\Bigg]\Bigg)\hat{\mu}_3^2+\Bigg(\frac{6835}{6804}-\frac{13670}{1701}\nu+\frac{27340}{1701}\nu^2\Bigg)\hat{\mu}_3^2\hat{\epsilon}_2^2
	\linechange
	+\Bigg(-\frac{1465}{486}+\frac{23230}{1701}\nu-\frac{10880}{1701}\nu^2 +\hat{\alpha_2}\Bigg[\frac{5}{9}-\frac{175}{81}\nu-\frac{20}{81}\nu^2\Bigg]+\hat{\alpha_{3}}\Bigg[\frac{200}{243}-\frac{100}{27}\nu+\frac{400}{243}\nu^2\Bigg]\Bigg)\hat{\epsilon}_2^2
	 +\Bigg(\frac{5}{243}-\frac{40}{243}\nu+\frac{80}{243}\nu^2\Bigg)\hat{\epsilon}_2^4
	 \linechange
	 +\Bigg(\frac{1600}{567}\nu-\frac{1600}{189}\nu^2\Bigg)\hat{\epsilon}_3^2\Bigg]\chi_s
	+\Bigg[-\frac{64160}{567}+\frac{17440}{567}\nu+\hat{\alpha_2} \Bigg(\frac{160}{9}+\frac{160}{81}\nu\Bigg)-\hat{\alpha_{3}} \Bigg(\frac{85600}{567}-\frac{44000}{567}\nu\Bigg)-\hat{\alpha_{5}} \Bigg(\frac{1120}{9}-\frac{4340}{81}\nu\Bigg)
	\linechange
	+\Bigg(\frac{13670}{1701}-\frac{23930}{1701}\nu+ \hat{\alpha_{3}}\Bigg[\frac{68350}{1701}-\frac{273400}{1701}\nu\Bigg] \Bigg)\hat{\mu}_3^2 +\Bigg(\frac{6835}{6804}-\frac{6835}{1701}\nu \Bigg)\hat{\mu}_3^2\hat{\epsilon}_2^2 +\Bigg(-\frac{1465}{486}+\frac{4520}{1701}\nu+\hat{\alpha_2} \Bigg[\frac{5}{9}+\frac{5}{81}\nu\Bigg] 
	\linechange
	+\hat{\alpha_{3}} \Bigg[\frac{200}{243}-\frac{800}{243}\nu\Bigg]\Bigg)\hat{\epsilon}_2^2+ \Bigg(\frac{5}{243}-\frac{20}{243}\nu\Bigg)\hat{\epsilon}_2^4
	\Bigg]\delta\chi_a\Bigg\}
	+\pi v^6\Bigg\{
	\Bigg[\frac{640}{3}-\frac{640}{3}\nu +\hat{\alpha_{3}}\Bigg(\frac{1600}{3}-\frac{800}{3}\nu\Bigg) +(10-40\nu)\hat{\epsilon}_2^2\Bigg]\chi_s
	\linechange
	+\Bigg[\frac{640}{3}+10\hat{\epsilon}_2^2 +\frac{1600}{3}\hat{\alpha_{3}} \Bigg]\delta\chi_a
	\Bigg\}
	+v^7\Bigg\{ \Bigg[-\frac{175520}{63}+\frac{7871090}{1323}\nu-\frac{4100}{3}\nu^2-\frac{199520}{1323}\nu^3+\hat{\alpha_2}\Bigg(\frac{16040}{21}-\frac{195280}{189}\nu
	\linechange
	- \frac{11600}{189}\nu^2
	+\frac{440}{63}\nu^3\Bigg) +\hat{\alpha_{3}}\Bigg(-\frac{11825200}{3969}+\frac{11267500}{3969}\nu - \frac{1322350}{1323}\nu^2+\frac{644800}{3969}\nu^3\Bigg) +\hat{\alpha_{4}}\Bigg(540-920\nu +\frac{1160}{3}\nu^2
	\linechange
	-\frac{20}{3}\nu^3\Bigg) +\hat{\alpha_{5}}\Bigg(-\frac{8560}{3}+\frac{169070}{27}\nu - \frac{68690}{27}\nu^2+\frac{1100}{27}\nu^3\Bigg) +\hat{\alpha_{7}}\Bigg(-2430+\frac{16785}{2}\nu - 2580\nu^2-15\nu^3\Bigg) +\hat{\mu}_3^2\Bigg(\frac{58105}{189}
	\linechange
	-\frac{22900195}{15876}\nu+\frac{8056835}{10584}\nu^2 +\frac{2844815}{7938}\nu^3\Bigg) +\hat{\mu}_3^2\hat{\alpha_2}\Bigg(-\frac{6835}{126}+\frac{127285}{567}\nu-\frac{32335}{1134}\nu^2-\frac{3410}{567}\nu^3\Bigg)
	+\hat{\mu}_3^2\hat{\alpha_{3}}\Bigg(\frac{524075}{1323}
	\linechange
	-\frac{5309275}{2646}\nu+\frac{2381500}{1323}\nu^2-\frac{592600}{1323}\nu^3\Bigg)
	+\hat{\mu}_3^2\hat{\alpha_{5}}\Bigg(\frac{6835}{9}-\frac{2795515}{648}\nu +\frac{20505}{4}\nu^2-\frac{6835}{81}\nu^3\Bigg)
	+\hat{\mu}_3^2\hat{\epsilon}_2^2\Bigg(\frac{3260435}{127008}
	\linechange
	-\frac{7054105}{31752}\nu +\frac{4326905}{7938}\nu^2-\frac{355490}{1323}\nu^3\Bigg)
	+\hat{\mu}_3^2\hat{\epsilon}_2^2\hat{\alpha_{2}}\Bigg(-\frac{6835}{1008} +\frac{485285}{9072}\nu-\frac{116195}{1134}\nu^2-\frac{6835}{567}\nu^3\Bigg)
	+\hat{\mu}_3^2\hat{\epsilon}_2^2\hat{\alpha_{3}}\Bigg(-\frac{34175}{3402} 
	\linechange
	+\frac{580975}{6804}\nu-\frac{341750}{1701}\nu^2+\frac{136700}{1701}\nu^3\Bigg)
	+\hat{\mu}_3^2\hat{\epsilon}_2^4\Bigg(-\frac{6835}{18144}+\frac{6835}{1512}\nu -\frac{6835}{378}\nu^2+\frac{13670}{567}\nu^3\Bigg)
	+\hat{\mu}_3^2\hat{\epsilon}_3^2\Bigg(-\frac{136700}{3969}\nu+\frac{136700}{567}\nu^2
	\linechange
	- \frac{546800}{1323}\nu^3\Bigg) +\hat{\mu}_3^4\Bigg(-\frac{129865}{1764}\nu +\frac{259730}{441}\nu^2-\frac{519460}{441}\nu^3\Bigg) +\hat{\mu}_3^4\hat{\epsilon}_2^2\Bigg(-\frac{9343445}{1016064}+\frac{9343445}{84672}\nu -\frac{9343445}{21168}\nu^2+\frac{9343445}{15876}\nu^3\Bigg) 
	\linechange
	+\hat{\mu}_3^4\hat{\alpha_{3}}\Bigg(-\frac{46717225}{190512} +\frac{794192825}{381024}\nu -\frac{233586125}{47628}\nu^2+\frac{46717225}{23814}\nu^3\Bigg) +\hat{\mu}_4^2\Bigg(\frac{289760}{11907}\nu-\frac{205600}{567}\nu^2 +\frac{1149440}{1323}\nu^3\Bigg) 
	\linechange
	+ \hat{\mu}_4^2\hat{\alpha_{3}}\Bigg( \frac{3586000}{11907}-\frac{23309000}{11907}\nu+\frac{14344000}{3969}\nu^2 -\frac{1793000}{1323}\nu^3\Bigg) + \hat{\mu}_4^2\hat{\epsilon}_2^2\Bigg(\frac{89650}{11907}-\frac{896500}{11907}\nu +\frac{986150}{3969}\nu^2-\frac{358600}{1323}\nu^3\Bigg) 
	\linechange
	+ \hat{\epsilon}_2^2\Bigg(-\frac{1193245}{15876}+\frac{938855}{2646}\nu-\frac{7469165}{31752}\nu^2 +\frac{83995}{1134}\nu^3\Bigg)+\hat{\epsilon}_2^2\hat{\alpha_2}\Bigg(\frac{1465}{72}-\frac{407885}{4536}\nu +\frac{5335}{162}\nu^2+\frac{2720}{567}\nu^3\Bigg) +\hat{\epsilon}_2^2\hat{\alpha_{3}}\Bigg(\frac{18850}{567} 
	\linechange
	-\frac{92825}{567}\nu+\frac{73700}{567}\nu^2-\frac{16000}{567}\nu^3\Bigg) +\hat{\epsilon}_2^2\hat{\alpha_4}\Bigg(\frac{135}{8}-\frac{635}{8}\nu +\frac{1145}{24}\nu^2-\frac{5}{6}\nu^3\Bigg) +\hat{\epsilon}_2^2\hat{\alpha_{5}}\Bigg(\frac{140}{9}-\frac{14315}{162}\nu +105\nu^2-\frac{140}{81}\nu^3\Bigg) 
	\linechange
	+\hat{\epsilon}_2^2\hat{\epsilon}_3^2\Bigg(\frac{50}{189}-\frac{1900}{567}\nu +\frac{7750}{567}\nu^2-\frac{3400}{189}\nu^3\Bigg) +\hat{\epsilon}_2^4\Bigg(\frac{1745}{1512}-\frac{1585}{162}\nu +\frac{12970}{567}\nu^2 -\frac{5000}{567}\nu^3\Bigg) +\hat{\epsilon}_2^4\hat{\alpha_2}\Bigg(-\frac{5}{36}+\frac{355}{324}\nu -\frac{170}{81}\nu^2
	\linechange
	-\frac{20}{81}\nu^3\Bigg) +\hat{\epsilon}_2^4\hat{\alpha_{3}}\Bigg(-\frac{25}{243} +\frac{425}{486}\nu-\frac{500}{243}\nu^2+\frac{200}{243}\nu^3\Bigg) +\hat{\epsilon}_2^6\Bigg(-\frac{5}{1296}+\frac{5}{108}\nu-\frac{5}{27}\nu^2 +\frac{20}{81}\nu^3\Bigg) +\hat{\epsilon}_3^2\Bigg(\frac{258520}{3969}\nu -\frac{966200}{3969}\nu^2
	\linechange
	+\frac{612800}{3969}\nu^3\Bigg) +\hat{\epsilon}_3^2\hat{\alpha_2}\Bigg( -\frac{400}{21}\nu +\frac{10400}{189}\nu^2+\frac{400}{63}\nu^3\Bigg) +\hat{\epsilon}_3^2\hat{\alpha_{3}}\Bigg(\frac{2000}{189}-\frac{13000}{189}\nu +\frac{8000}{63}\nu^2-\frac{1000}{21}\nu^3\Bigg) +\hat{\epsilon}_4^2\Bigg(\frac{28705}{1764}\nu
	\linechange
	-\frac{28705}{294}\nu^2 +\frac{57410}{441}\nu^3\Bigg) \Bigg]\chi_s
	+\Bigg[-\frac{175520}{63} +\frac{3039410}{1323}\nu+\frac{29300}{1323}\nu^2 +\hat{\alpha_2}\Bigg(\frac{16040}{21}-\frac{23200}{189}\nu-\frac{4360}{189}\nu^2\Bigg)
	\linechange
	+\hat{\alpha_{3}}\Bigg(-\frac{11825200}{3969}+\frac{5354900}{3969}\nu -\frac{1289600}{3969}\nu^2\Bigg)
	+\hat{\alpha_4}\Bigg(540-380\nu+\frac{20}{3}\nu^2\Bigg) +\hat{\alpha_{5}}\Bigg(-\frac{8560}{3}+\frac{72770}{27}\nu -\frac{17050}{27}\nu^2\Bigg)
	\linechange
	+\hat{\alpha_{7}}\Bigg(-2430+\frac{9495}{2}\nu -105\nu^2\Bigg)
	+\hat{\mu}_3^2\Bigg(\frac{58105}{189}-\frac{15599425}{15876}\nu-\frac{5525}{24}\nu^2\Bigg) +\hat{\mu}_3^2\hat{\alpha_2}\Bigg(-\frac{6835}{126}+\frac{50425}{567}\nu +\frac{11965}{1134}\nu^2\Bigg) 
	\linechange
	+\hat{\mu}_3^2\hat{\alpha_{3}}\Bigg(\frac{524075}{1323}-\frac{341800}{189}\nu +\frac{1185200}{1323}\nu^2\Bigg)
	+\hat{\mu}_3^2\hat{\alpha_{5}}\Bigg(\frac{6835}{9}-\frac{2180365}{648}\nu +\frac{211885}{162}\nu^2\Bigg) +\hat{\mu}_3^2\hat{\epsilon}_2^2\Bigg(\frac{3260435}{127008}-\frac{2245385}{15876}\nu 
	\linechange
	+\frac{1230335}{7938}\nu^2\Bigg)
	+\hat{\mu}_3^2\hat{\epsilon}_2^2\hat{\alpha_{2}}\Bigg(-\frac{6835}{1008} +\frac{34175}{1296}\nu+\frac{6835}{2268}\nu^2\Bigg)
	+\hat{\mu}_3^2\hat{\epsilon}_2^2\hat{\alpha_{3}}\Bigg(-\frac{34175}{3402} +\frac{136700}{1701}\nu
	-\frac{273400}{1701}\nu^2\Bigg)
	+\hat{\mu}_3^2\hat{\epsilon}_2^4\Bigg(-\frac{6835}{18144}
	\linechange
	+\frac{6835}{2268}\nu -\frac{6835}{1134}\nu^2\Bigg)
	+\hat{\mu}_3^4\Bigg(-\frac{7005875}{31752}\nu+\frac{7005875}{7938}\nu^2\Bigg)
	+\hat{\mu}_3^4\hat{\alpha_{3}}\Bigg(-\frac{46717225}{190512} +\frac{46717225}{23814}\nu
	-\frac{46717225}{11907}\nu^2\Bigg)
	\linechange
	+\hat{\mu}_3^4\hat{\epsilon}_2^2\Bigg(-\frac{9343445}{1016064}+\frac{9343445}{127008}\nu -\frac{9343445}{63504}\nu^2\Bigg) 
	+ \hat{\mu}_4^2\Bigg(\frac{31840}{147}\nu-\frac{31840}{49}\nu^2\Bigg)
	+ \hat{\mu}_4^2\hat{\alpha_{3}}\Bigg(\frac{3586000}{11907} -\frac{7172000}{3969}\nu
	\linechange
	+\frac{3586000}{1323}\nu^2\Bigg)
	+ \hat{\mu}_4^2\hat{\epsilon}_2^2\Bigg(\frac{89650}{11907}-\frac{179300}{3969}\nu +\frac{89650}{1323}\nu^2\Bigg)
	+\hat{\epsilon}_2^2\Bigg(-\frac{1193245}{15876}+\frac{65255}{882}\nu -\frac{1742365}{31752}\nu^2\Bigg)
	+\hat{\epsilon}_2^2\hat{\alpha_2}\Bigg(\frac{1465}{72}
	\linechange
	-\frac{71105}{4536}\nu
	-\frac{1130}{567}\nu^2\Bigg)
	+\hat{\epsilon}_2^2\hat{\alpha_{3}}\Bigg(\frac{18850}{567} -\frac{27800}{189}\nu+\frac{32000}{567}\nu^2\Bigg)
	+\hat{\epsilon}_2^2\hat{\alpha_4}\Bigg(\frac{135}{8}-\frac{95}{8}\nu +\frac{5}{24}\nu^2\Bigg)
	+\hat{\epsilon}_2^2\hat{\alpha_{5}}\Bigg(\frac{140}{9}-\frac{11165}{162}\nu 
	\linechange
	+\frac{2170}{81}\nu^2\Bigg)
	+\hat{\epsilon}_2^2\hat{\epsilon}_3^2\Bigg(\frac{50}{189}-\frac{100}{63}\nu +\frac{50}{21}\nu^2\Bigg) 
	+\hat{\epsilon}_2^4\Bigg(\frac{1745}{1512}-\frac{50}{9}\nu+\frac{710}{189}\nu^2\Bigg)
	+\hat{\epsilon}_2^4\hat{\alpha_2}\Bigg(-\frac{5}{36}+\frac{175}{324}\nu +\frac{5}{81}\nu^2\Bigg)
	\linechange
	+\hat{\epsilon}_2^4\hat{\alpha_{3}}\Bigg(-\frac{25}{243}+\frac{200}{243}\nu
	-\frac{400}{243}\nu^2\Bigg)
	-\hat{\epsilon}_2^6\Bigg(\frac{5}{1296}-\frac{5}{162}\nu+\frac{5}{81}\nu^2\Bigg)
	+\hat{\epsilon}_3^2\Bigg(\frac{3800}{189}\nu-\frac{3800}{63}\nu^2\Bigg)
	+\hat{\epsilon}_3^2\hat{\alpha_{3}}\Bigg(\frac{2000}{189}-\frac{4000}{63}\nu
	\linechange
	+\frac{2000}{21}\nu^2\Bigg)
	-\hat{\epsilon}_4^2\Bigg(\frac{28705}{1764}\nu-\frac{28705}{882}\nu^2\Bigg)
	\Bigg]\delta\chi_a
	\Bigg\}\label{phase_SO}
	\end{eqnarray}
\end{widetext}

\begin{widetext}
	\begin{eqnarray}
	\psi_{\rm SS}(f)&&=v^4\Bigg\{
	\Bigg[-10\kappa_+ -\frac{5}{8}\hat{\epsilon}_2^2-15\kappa_+\hat{\alpha_4} -\delta\kappa_-\Bigg(10+15\hat{\alpha_4}\Bigg) +\Big(-40+20\kappa_++\frac{5}{2}\hat{\epsilon}_2^2-\hat{\alpha_4}[60-30\kappa_+]\Big)\nu \Bigg]\chi_s^2 
	+\Bigg[-20\kappa_-
	\linechange
	-30\kappa_-\hat{\alpha_4} -\delta\Bigg(20\kappa_++30\kappa_+\hat{\alpha_4}+\frac{5}{4}\hat{\epsilon}_2^2\Bigg) +\nu\kappa_-\Bigg(40+60\hat{\alpha_4}\Bigg)\Bigg]\chi_s\chi_a 
	+\Bigg[-10\kappa_+ -\frac{5}{8}\hat{\epsilon}_2^2-15\kappa_+\hat{\alpha_4} -\delta\kappa_-\Bigg(10+15\hat{\alpha_4}\Bigg)
	\linechange
	+\Bigg(40+20\kappa_+ +\hat{\alpha_4}[60 +30\kappa_+]
	\Bigg)\nu\Bigg]\chi_a^2
	\Bigg\}	
	+v^6\Bigg\{\Bigg[-\frac{1120}{9}+\frac{1150}{7}\kappa_++\kappa_-\delta\Bigg(\frac{1150}{7} -\frac{690}{7}\nu\Bigg) +\Bigg(\frac{38600}{63}-\frac{2990}{7}\kappa_+\Bigg)\nu 
	\linechange
	-\Bigg(\frac{3880}{21}-\frac{1940}{21}\kappa_+\Bigg)\nu^2 +\frac{1600}{63}\nu^2\hat{\epsilon}_3^2 +\hat{\alpha_2}\Bigg(-30\kappa_+ -\Bigg[30+\frac{10}{3}\nu\Bigg]\delta\kappa_- +\Bigg[-120+\frac{170}{3}\kappa_+\Bigg]\nu +\Bigg[-\frac{40}{3}+\frac{20}{3}\kappa_+\Bigg]\nu^2\Bigg)
	\linechange
	-\hat{\alpha_{3}}\Bigg(\frac{3200}{9}-\frac{1600}{3}\nu+\frac{1600}{9}\nu^2\Bigg)
	+\hat{\alpha_4}\Bigg(\frac{1070}{7}\kappa_+ +\Bigg[\frac{1070}{7}-\frac{550}{7}\nu\Bigg]\delta\kappa_-
	+\Bigg[\frac{4280}{7}-\frac{2690}{7}\kappa_+\Bigg]\nu
	-\Bigg[\frac{2200}{7}-\frac{1100}{7}\kappa_+\Bigg]\nu^2
	\Bigg)
	\linechange
	+\hat{\alpha_6}\Bigg(-\frac{1600}{9}+\frac{700}{3}\kappa_+ 
	+\Bigg[\frac{700}{3}-\frac{500}{3}\nu\Bigg]\delta\kappa_-
	+\Bigg[\frac{1600}{9} -\frac{1900}{3}\kappa_+\Bigg]\nu
	+\Bigg[\frac{2000}{9}+\frac{200}{3}\kappa_+\Bigg]\nu^2
	\Bigg) +\hat{\mu}_3^2\Bigg(-\frac{95}{7}\kappa_+ +\Bigg[-\frac{95}{7}
	\linechange
	+\frac{13675}{252}\nu\Bigg]\delta\kappa_- +\Bigg[-\frac{6835}{126}+\frac{20515}{252}\kappa_+\Bigg]\nu +\Bigg[\frac{13670}{63}-\frac{6835}{63}\kappa_+\Bigg]\nu^2\Bigg) +\hat{\mu}_3^2\hat{\alpha_4}\Bigg(-\frac{6835}{168}\kappa_+ -\Bigg[\frac{6835}{168}-\frac{6835}{42}\nu\Bigg]\delta\kappa_- 
	\linechange
	-\Bigg[\frac{6835}{42}-\frac{6835}{28}\kappa_+\Bigg]\nu+\Bigg[\frac{13670}{21}-\frac{6835}{21}\kappa_+\Bigg]\nu^2\Bigg)
	-\hat{\mu}_3^2\hat{\epsilon}_2^2\Bigg(\frac{6835}{2016}-\frac{6835}{252}\nu +\frac{6835}{126}\nu^2\Bigg) +\hat{\epsilon}_2^2\Bigg( \frac{470}{63}-\frac{5}{6}\kappa_+ -\Bigg[\frac{5}{6}-\frac{20}{9}\nu\Bigg]\delta\kappa_- 
	\linechange
	-\Bigg[\frac{730}{21}-\frac{35}{9}\kappa_+\Bigg]\nu +\Bigg[\frac{1240}{63}-\frac{20}{9}\kappa_+\Bigg]\nu^2\Bigg)
	-\hat{\epsilon}_2^2\hat{\alpha_2}\Bigg(\frac{15}{8}-\frac{175}{24}\nu -\frac{5}{6}\nu^2\Bigg)
	-\hat{\epsilon}_2^2\hat{\alpha_{3}}\Bigg(\frac{100}{9}- 50\nu +\frac{200}{9}\nu^2\Bigg)
	+\hat{\epsilon}_2^2\hat{\alpha_4}\Bigg(-\frac{5}{6}\kappa_+ 
	\linechange
	-\Bigg[\frac{5}{6}-\frac{10}{3}\nu\Bigg]\delta\kappa_- -\Bigg[\frac{10}{3}-5\kappa_+\Bigg]\nu +\Bigg[\frac{40}{3}- \frac{20}{3}\kappa_+\Bigg]\nu^2\Bigg) -\hat{\epsilon}_2^4\Bigg(\frac{5}{24}-\frac{5}{3}\nu+\frac{10}{3}\nu^2\Bigg)
	\Bigg]\chi_s^2
	+\Bigg[ \Bigg(\frac{2300}{7}-\frac{5980}{7}\nu +\frac{3880}{21}\nu^2\Bigg)\kappa_- 
	\linechange
	+	\Bigg(-\frac{2240}{9}+\frac{2300}{7}\kappa_+ +\Bigg(\frac{560}{9}-\frac{1380}{7}\kappa_+\Bigg)\nu\Bigg)\delta
	+\hat{\alpha_2}\Bigg(\Bigg[-60+\frac{340}{3}\nu+\frac{40}{3}\nu^2\Bigg]\kappa_- - \Bigg(60+\frac{20}{3}\nu\Bigg)\delta\kappa_+\Bigg) -\hat{\alpha_{3}}\delta\Bigg(\frac{6400}{9}-\frac{1600}{3}\nu\Bigg)
	\linechange
	+\hat{\alpha_4}\Bigg(\Bigg[\frac{2140}{7}-\frac{5380}{7}\nu +\frac{2200}{7}\nu^2\Bigg]\kappa_-+\Bigg[\frac{2140}{7}-\frac{1100}{7}\nu\Bigg]\kappa_+\delta\Bigg) +\hat{\alpha_6}\Bigg(\Bigg[\frac{1400}{3}-\frac{3800}{3}\nu +\frac{400}{3}\nu^2\Bigg]\kappa_- +\Bigg[-\frac{3200}{9}+\frac{1400}{3}\kappa_+ 
	\linechange
	-\Bigg(\frac{5600}{9}+\frac{1000}{3}\kappa_+\Bigg)\nu\Bigg]\delta\Bigg)
	+\hat{\mu}_3^2\Bigg(\Bigg[-\frac{190}{7}+\frac{20515}{126}\nu-\frac{13670}{63}\nu^2 \Bigg]\kappa_- +\Bigg[-\frac{190}{7}+\frac{13675}{126}\nu\Bigg]\delta\kappa_+\Bigg) +\hat{\mu}_3^2\hat{\alpha_4}\Bigg(\Bigg[-\frac{6835}{84}+\frac{6835}{14}\nu 
	\linechange
	-\frac{13670}{21}\nu^2\Bigg]\kappa_-  -\Bigg[\frac{6835}{84}-\frac{6835}{21}\nu\Bigg] \delta\kappa_+\Bigg) - \hat{\mu}_3^2\hat{\epsilon}_2^2\delta\Bigg(\frac{6835}{1008}-\frac{6835}{252}\nu\Bigg) +\hat{\epsilon}_2^2\Bigg(\Bigg[-\frac{5}{3}+\frac{70}{9}\nu- \frac{40}{9}\nu^2\Bigg]\kappa_- +\Bigg[\frac{940}{63}-\frac{5}{3}\kappa_+ -\Bigg(\frac{650}{63}
	\linechange
	-\frac{40}{9}\kappa_+\Bigg)\nu\Bigg]\delta\Bigg) - \hat{\epsilon}_2^2\hat{\alpha_2}\delta\Bigg(\frac{15}{4}+\frac{5}{12}\nu\Bigg)
	-\hat{\epsilon}_2^2\hat{\alpha_{3}}\delta\Bigg(\frac{200}{9}-50\nu\Bigg) 
	+\hat{\epsilon}_2^2\hat{\alpha_{4}}\Bigg(\Bigg[-\frac{5}{3}+10\nu -\frac{40}{3}\nu^2\Bigg]\kappa_-  +\Bigg[-\frac{5}{3}+\frac{20}{3}\nu\Bigg]\delta\kappa_+\Bigg)
	\linechange
	 - \hat{\epsilon}_2^4\delta\Bigg(\frac{5}{12}-\frac{5}{3}\nu \Bigg)
	\Bigg]\chi_s\chi_a +\Bigg[-\frac{1120}{9}+\frac{1150}{7}\kappa_+ -\Bigg(\frac{3320}{63} +\frac{2990}{7}\kappa_+\Bigg)\nu +\Bigg(\frac{3880}{21}+\frac{1940}{21}\kappa_+\Bigg)\nu^2 +\Bigg(\frac{1150}{7}-\frac{690}{7}\nu\Bigg)\delta\kappa_-  
	\linechange
	+\hat{\alpha_2}\Bigg(-30\kappa_+ +\Bigg[120+\frac{170}{3}\kappa_+\Bigg]\nu +\Bigg[\frac{40}{3}+\frac{20}{3}\kappa_+\Bigg]\nu^2 -\Bigg[30+\frac{10}{3}\nu\Bigg]\delta\kappa_-\Bigg) - \hat{\alpha_{3}}\Bigg(\frac{3200}{9}-\frac{12800}{9}\nu\Bigg) + \hat{\alpha_{4}}\Bigg( \frac{1070}{7}\kappa_+ 
	\linechange
	-\Bigg[\frac{4280}{7}+\frac{2690}{7}\kappa_+\Bigg]\nu +\Bigg[\frac{2200}{7}+\frac{1100}{7}\kappa_+\Bigg]\nu^2 +\Bigg[\frac{1070}{7}-\frac{550}{7}\nu\Bigg]\delta\kappa_-\Bigg) + \hat{\alpha_{6}}\Bigg(
	-\frac{1600}{9}+\frac{700}{3}\kappa_+ -\Bigg[\frac{800}{9}+\frac{1900}{3}\kappa_+\Bigg]\nu 
	\linechange
	+ \Bigg[\frac{400}{3}+\frac{200}{3}\kappa_+ \Bigg]\nu^2 +\Bigg[\frac{700}{3}-\frac{500}{3}\nu\Bigg]\delta\kappa_-\Bigg) 
	+\hat{\mu}_3^2 \Bigg(-\frac{95}{7}\kappa_+ + \Bigg[\frac{6835}{126}+\frac{20515}{252} \kappa_+\Bigg]\nu -\Bigg[\frac{13670}{63}+\frac{6835}{63} \kappa_+ \Bigg]\nu^2  
	\linechange
	-\Bigg[\frac{95}{7}-\frac{13675}{252}\nu\Bigg]\delta\kappa_-\Bigg) +\hat{\mu}_3^2\hat{\alpha_4} \Bigg(-\frac{6835}{168}\kappa_+ + \Bigg[\frac{6835}{42}+\frac{6835}{28}\kappa_+\Bigg]\nu -\Bigg[\frac{13670}{21} +\frac{6835}{21}\kappa_+\Bigg]\nu^2 -\Bigg[\frac{6835}{168}-\frac{6835}{42}\nu\Bigg]\delta\kappa_-\Bigg) 
	\linechange
	+\hat{\mu}_3^2\hat{\epsilon}_2^2 \Bigg(-\frac{6835}{2016}+\frac{6835}{504}\nu\Bigg) +\hat{\epsilon}_2^2\Bigg(\frac{470}{63}-\frac{5}{6}\kappa_+ -\Bigg[\frac{340}{63}-\frac{35}{9}\kappa_+\Bigg]\nu -\Bigg[\frac{40}{9} + \frac{20}{9}\kappa_+\Bigg]\nu^2 -\Bigg[\frac{5}{6}-\frac{20}{9}\nu\Bigg]\delta\kappa_-\Bigg) 
	-\hat{\epsilon}_2^2\hat{\alpha_2}\Bigg(\frac{15}{8}+\frac{5}{24}\nu\Bigg)
	\linechange
	-\hat{\epsilon}_2^2\hat{\alpha_{3}}\Bigg(\frac{100}{9}-\frac{400}{9}\nu\Bigg) +\hat{\epsilon}_2^2\hat{\alpha_4}\Bigg(-\frac{5}{6}\kappa_+ +\Bigg[\frac{10}{3}+5\kappa_+\Bigg]\nu  -\Bigg[\frac{40}{3}+\frac{20}{3}\kappa_+ \Bigg]\nu^2 -\Bigg[\frac{5}{6}-\frac{10}{3}\nu\Bigg]\delta\kappa_-\Bigg) -\hat{\epsilon}_2^4\Bigg(\frac{5}{24}-\frac{5}{6}\nu\Bigg) 
	\Bigg]\chi_a^2
	\Bigg\}
	\label{phase_SS}
	\end{eqnarray}
\end{widetext}

As a consistency check, we confirm the recovery of the corresponding GR expression for the TaylorF2 phasing for aligned spin binaries (see Refs.~\cite{MKAF2016,ABFO08, Wade2013}) in the limit, $\mu_2=\mu_3=\mu_4=\mu_5=\epsilon_2=\epsilon_3=\epsilon_4=\alpha_0=\alpha_2=\alpha_3=\alpha_4=\alpha_5=\alpha_6=\alpha_7=1$. 
We also update Table I of Ref.~\cite{Kastha2018} to explicitly show the appearances of the parameters $\mu_l$ and $\epsilon_l$ at various PN order of the phasing formula (see Table~\ref{Non-GR_Params_at_every_PN_order}).

\begin{center}
	\begin{table}
		\begin{tabular}{||c | c | c ||} 
			\hline
			PN order  & frequency dependences  & Multipole coefficients  \\ [0.5ex] 
			\hline\hline
			0 PN & $f^{-5/3}$ & $\mu_2$  \\  [1ex]
			\hline
			1 PN & $f^{-1}$  &  $\mu_2$, $\mu_3$, $\epsilon_2$ \\ [1ex]
			\hline
			1.5 PN & $f^{-2/3}$ &  $\mu_2$, \underline{$\epsilon_2$} \\ [1ex]
			\hline
			2 PN & $f^{-1/3}$ &  $\mu_2$, $\mu_3$, $\mu_4$, $\epsilon_2$, $\epsilon_3$  \\ [1ex]
			\hline
			2.5 PN log & $\log f$  & $\mu_2$, $\mu_3$, $\epsilon_2$, $\underline{\epsilon_3}$ \\ [1ex]
			\hline
			3 PN & $f^{1/3}$ & $\mu_2$, $\mu_3$, $\mu_4$, $\mu_5$, $\epsilon_2$, $\epsilon_3$, $\epsilon_4$  \\ [1ex]
			\hline
			3 PN log &$f^{1/3}\log f$& $\mu_2$  \\ [1ex]
			\hline
			3.5 PN & $f^{2/3}$ &  $\mu_2$, $\mu_3$, $\mu_4$, $\epsilon_2$, $\epsilon_3$, $\underline{\epsilon_4}$\\ [1ex] 
			\hline
		\end{tabular}
		\caption{Update of the summary given in Table I of Ref.~\cite{Kastha2018} for the multipolar structure of the PN phasing formula. Contribution of various multipoles to different phasing coefficients and their frequency dependences are tabulated. The additional multipole coefficients appearing due to spin are underlined. Following the definitions introduced in Ref.~\cite{Kastha2018}, $\mu_l$ refer to mass-type multipole moments and $\epsilon_l$ refer to current-type multipole moments.}
		\label{Non-GR_Params_at_every_PN_order}
	\end{table}
\end{center}
One of the salient features of the parametrized multipolar
spinning phasing derived here is the presence of $\epsilon_2$ at 1.5PN order 
and $\epsilon_3$ at 2.5PN order (logarithmic) due to the spin-orbit
interactions and hence not present in the non-spinning phasing. 
Though at 2PN and 3PN order, due to the spin-spin interactions,
there are no additional multipole moments compared to the non-spinning
systems, these are the orders at which $\kappa_{+,-}$ appear.
This has interesting interpretation as  $\kappa_{+,-}$ can be thought of
as parametrizing potential deviations from BH
nature~\cite{KAM2017,Krishnendu:2018nqa} as
binaries comprising of non-BHs will have $\kappa_{+,-}$ to be different
from 2 and 0, respectively, which are the unique values corresponding to
binary black holes.  The cross-terms of the multipole coefficients with
$\kappa_{+,-}$ showcase the degeneracy between binary black holes in
alternative theories and non-BHs in GR. As one can see from
Eq.~(\ref{phase_SS}), $\mu_2,\mu_3$ and $\epsilon_2$ are the multipole coefficients
which are sensitive to the non-BH nature (vis-a-vis the above mentioned
parametrization). As can be seen from the phasing formula, these
imprints will be higher order corrections to the multipole coefficients
and may not influence their estimates unless the values of
$\kappa_{+,-}$ are sufficiently high.

\begin{figure*}[t]
	\centering
	\includegraphics[scale=0.38]{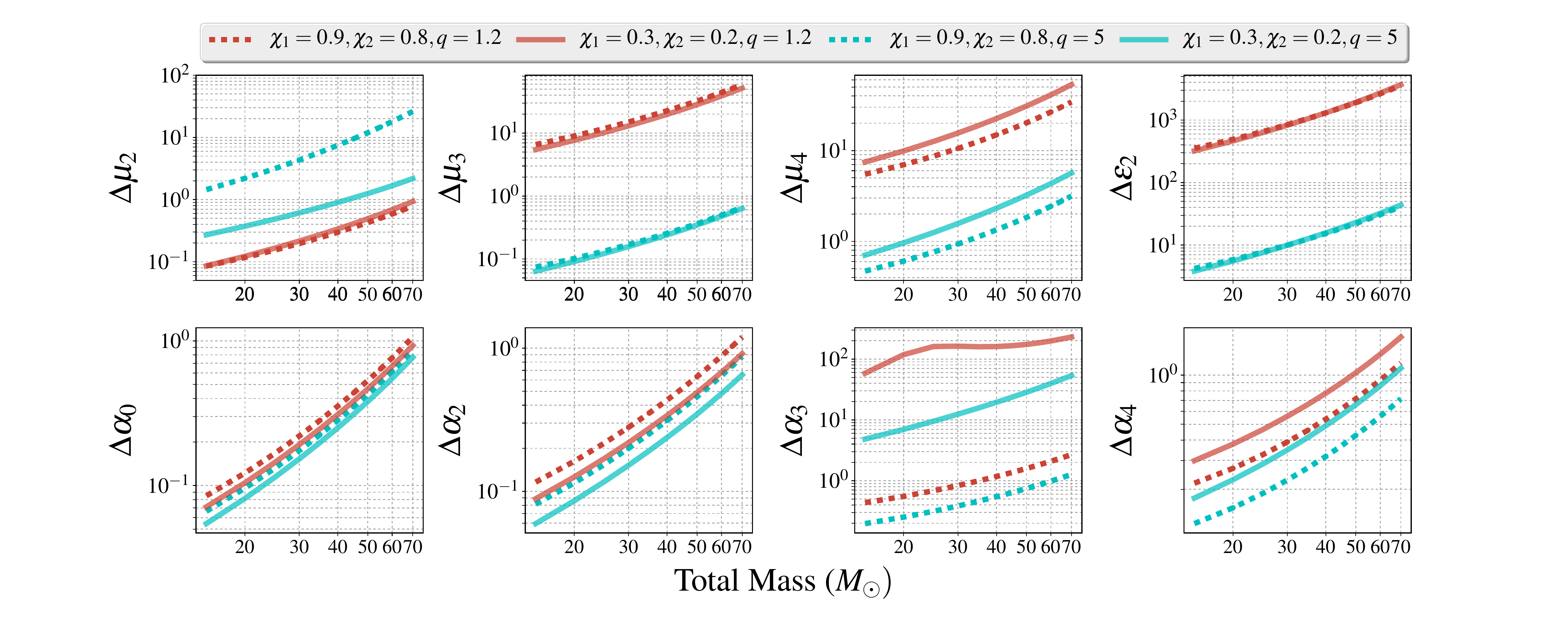}
	\caption{Projected $1\sigma$ errors on the multipole and the energy coefficients as a function of total mass for two different mass ratios $q=m_1/m_2=1.2,5$ and two spin configurations, $\chi_1=0.9, \chi_2=0.8$ and $\chi_1=0.3, \chi_2=0.2$ for the second generation detector network. All the sources are at a fixed luminosity distance of 100 Mpc with the angular position and orientations to be $\theta=\pi/6, \phi=\pi/3,\psi=\pi/6, \iota=\pi/5$. To obtain the numerical estimates showed in this plot, we also consider a prior distribution on $\phi_c$. To be precise, we assume the prior on $\phi_c$ for each detector in the network to follow a Gaussian distribution with a zero mean and a variance of $1/\pi^2.$}\label{fig:2G_network_bounds}
\end{figure*}

\section{Parameter Estimation Scheme}
\label{sec:pe}
In this section, we briefly describe the semi-analytical Fisher information matrix based parameter estimation scheme~\cite{Rao45,Cramer46,CF94,AISS05} used in our analysis. We also discuss the leading order  bounds on the systematics of the estimated parameters due to the difference between the spinning and non-spinning waveforms  in the Appendix for LISA.

For ${\vec \theta}$ being the set of parameters defining the GW signal $\tilde{h}(f;\vec{\theta})$, the Fisher information matrix is defined as
\begin{equation}
\label{eq:fisher}
\Gamma_{mn}=\Bigg\langle\frac{\partial\tilde{h}(f;\vec{\theta})}{\partial\theta_m},\frac{\partial\tilde{h}(f;\vec{\theta})}{\partial\theta_n}\Bigg\rangle,
\end{equation}
where $\langle...,...\rangle$ is the inner product weighted by the detector noise. To be precise,

\begin{equation}
\label{eq:innerproduct}
\langle a,b\rangle=2\int_{f_{\rm low}}^{\rm f_{\rm high}}\frac{a(f)\,b^*(f)+a^*(f)\, b(f)}{S_h(f)}\,df \,.
\end{equation}
Here $`*$' denotes the complex conjugation and $S_h(f)$ is the one-sided
noise power spectral density (PSD) of the detector while $f_{\rm low}$ and
$f_{\rm high}$ denote the lower and upper limits of the integration.
Though   $f_{\rm low}$ arises from the detector sensitivity, $f_{\rm high}$ is defined by the frequency at
the last stable orbit of the binary beyond which the PN approximation
would break down. In the large signal-to-noise ratio (SNR) limit, the
distribution of the inferred parameters follow a Gaussian distribution around their respective true values for which the variance-covariance matrix of the errors on the parameters is simply the inverse of the Fisher matrix, $$C^{mn}=(\Gamma^{-1})^{mn},$$ and the $1\sigma$ statistical error is, $\Delta_{\text{stat}}\theta_m=\sqrt{C^{mm}}$.

\begin{figure*}[t]
	\centering
	\includegraphics[scale=0.28]{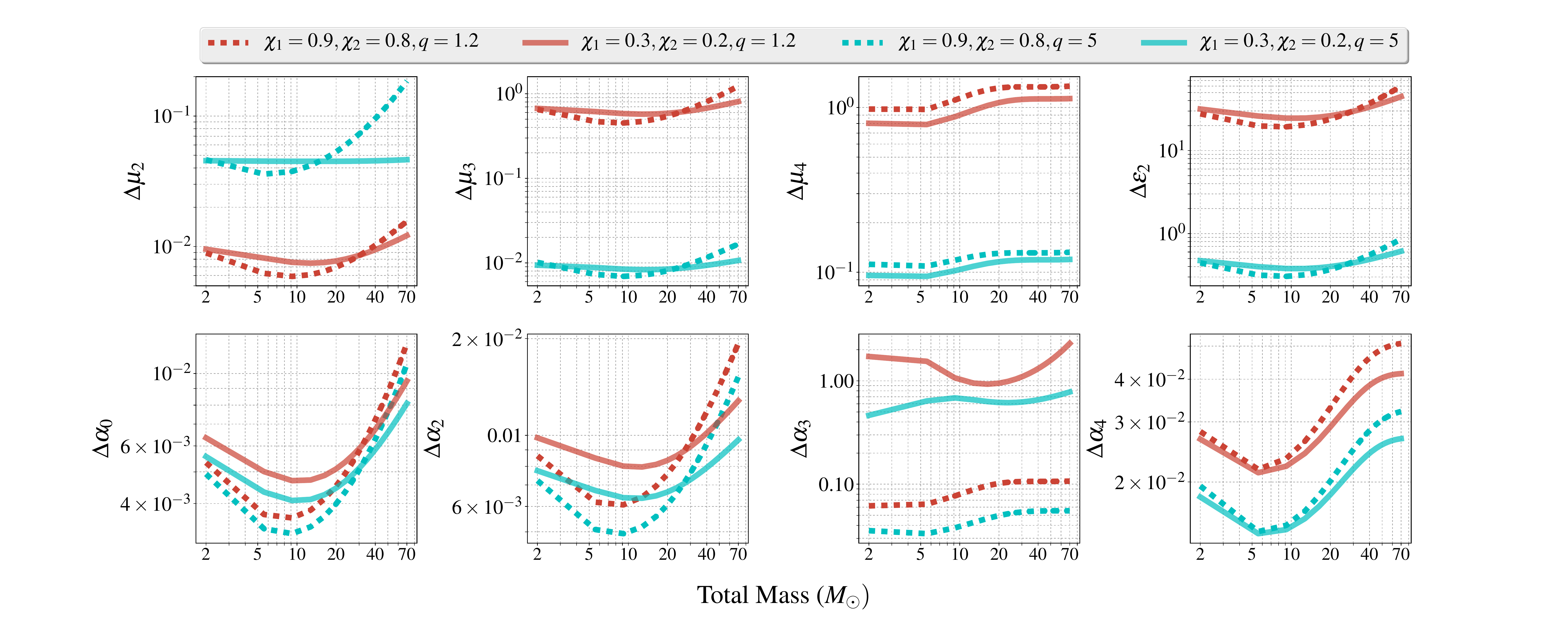}
	\caption{Projected $1\sigma$ errors on the multipole and the energy coefficients as a function of total mass for two different mass ratios $q=m_1/m_2=1.2,5$ and two spin configurations, $\chi_1=0.9, \chi_2=0.8$ and $\chi_1=0.3, \chi_2=0.2$ for the third generation detector network. All the sources are at a fixed luminosity distance of 100 Mpc with the angular position and orientations to be $\theta=\pi/5, \phi=\pi/6,\psi=\pi/4, \iota=\pi/4$. To obtain the numerical estimates showed in this plot, we also consider a prior distribution on $\phi_c$. To be precise, we assume the prior on $\phi_c$ for each detector in the network to follow a Gaussian distribution with a zero mean and a variance of $1/\pi^2.$}\label{fig:3G_network_bounds}
\end{figure*}

Fisher information matrix method, by default, assumes a flat prior
distribution in the range $[-\infty,\infty]$ on all the parameters to be
estimated~\cite{CF94,Vallisneri07}. In contrast, in the large SNR
limit, a Gaussian prior can also be implemented on the desired parameter as described
in Ref.~\cite{CF94}. For our purpose, we employ a Gaussian prior on $\phi_c$ centered 
around $\phi_c=0$ with a variance of about $\pi^2$. This choice is somewhat adhoc but 
ensures that the width of the Gaussian is not too small to significantly
influence the result  but helps us deal with the
ill-conditionedness of the Fisher matrix.
This also restricts the prior range to exceed to the unphysical domain beyond $\pm \pi$. 
Hence our modified Fisher matrix has the following form,

\begin{equation}
 \Gamma^\prime=\Gamma+\Gamma^{(0)},
\end{equation}
where $\Gamma^{(0)}$ is a diagonal matrix with only one non-zero element
corresponding to $\Gamma^{(0)}_{\phi_c\phi_c}$ component. We use this
modified Fisher matrix ($\Gamma^{\prime}$) for the estimation of
$1\sigma$ statistical errors which also can be interpreted as the
$1\sigma$ upper bounds on any deviation of these coefficients from GR
value.

We estimate the statistical errors on various multipole coefficients while considering an 
eight dimensional parameter space, 
$\{ t_c, \phi_c, \log \mathcal{A}, \log\mathcal{M}_c, \log\nu, \chi_s, \chi_a, \mu_\ell$ or $\epsilon_\ell$ or $\alpha_m\}$ to specify the true GW signal.

\begin{figure*}[t]
	\centering
	\includegraphics[scale=0.45]{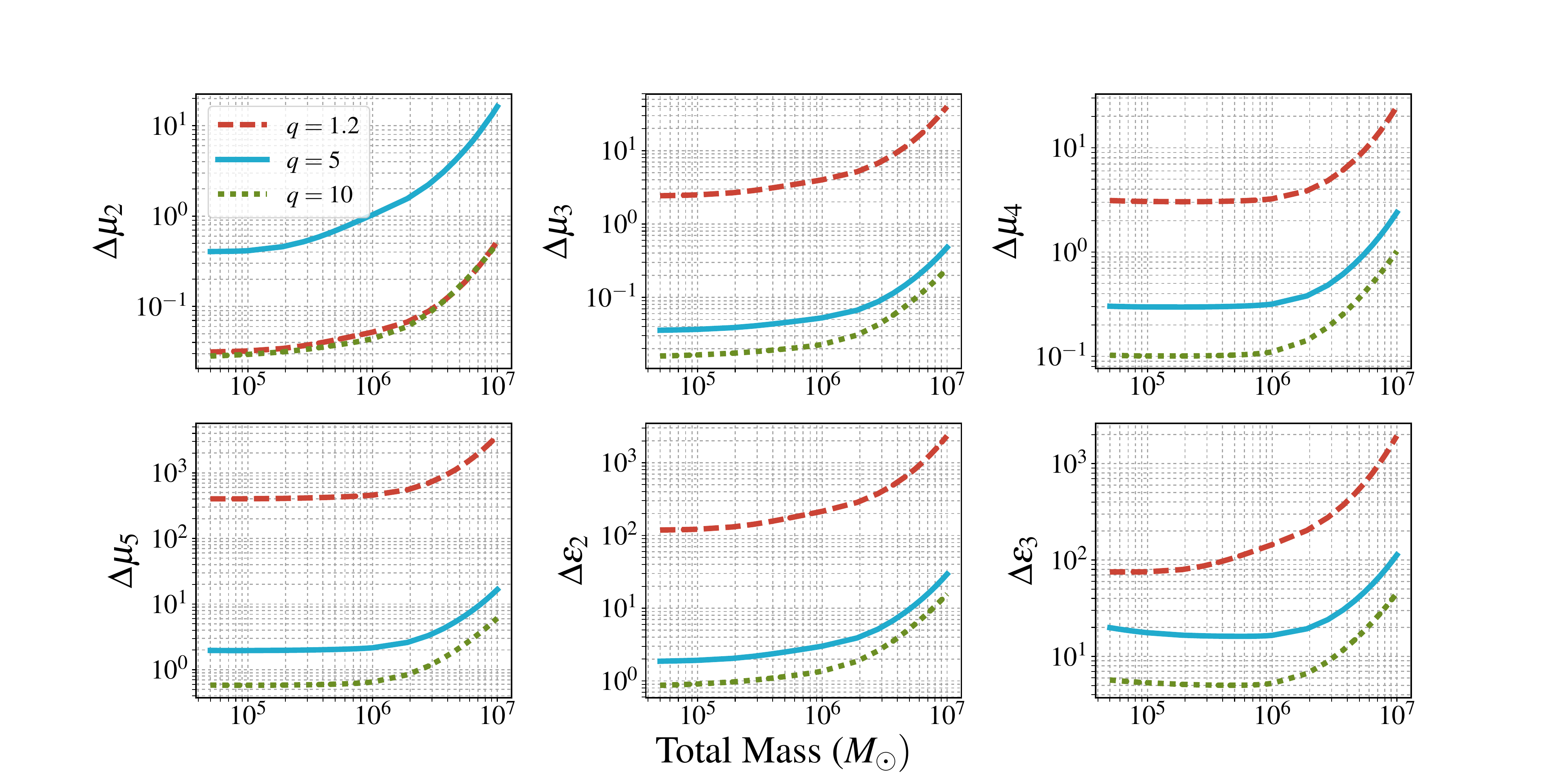}
	\caption{Projected $1\sigma$ errors on the multipole coefficients as a function of total mass for three different mass ratios $q=m_1/m_2=1.2,5$ and $10$ in case of LISA noise PSD. We assume $\chi_1=0.9, \chi_2=0.8$. All the sources are considered to be at a fixed luminosity distance of 3 Gpc.  To obtain the numerical estimates showed in this plot, we also consider a prior distribution on $\phi_c$. To be precise, we assume $\phi_c$ to follow a gaussian distribution with a zero mean and a variance of $1/\pi^2.$ }\label{fig:sys_mu_highspin}
\end{figure*}

\begin{figure*}[t]
	\centering
	\includegraphics[scale=0.45]{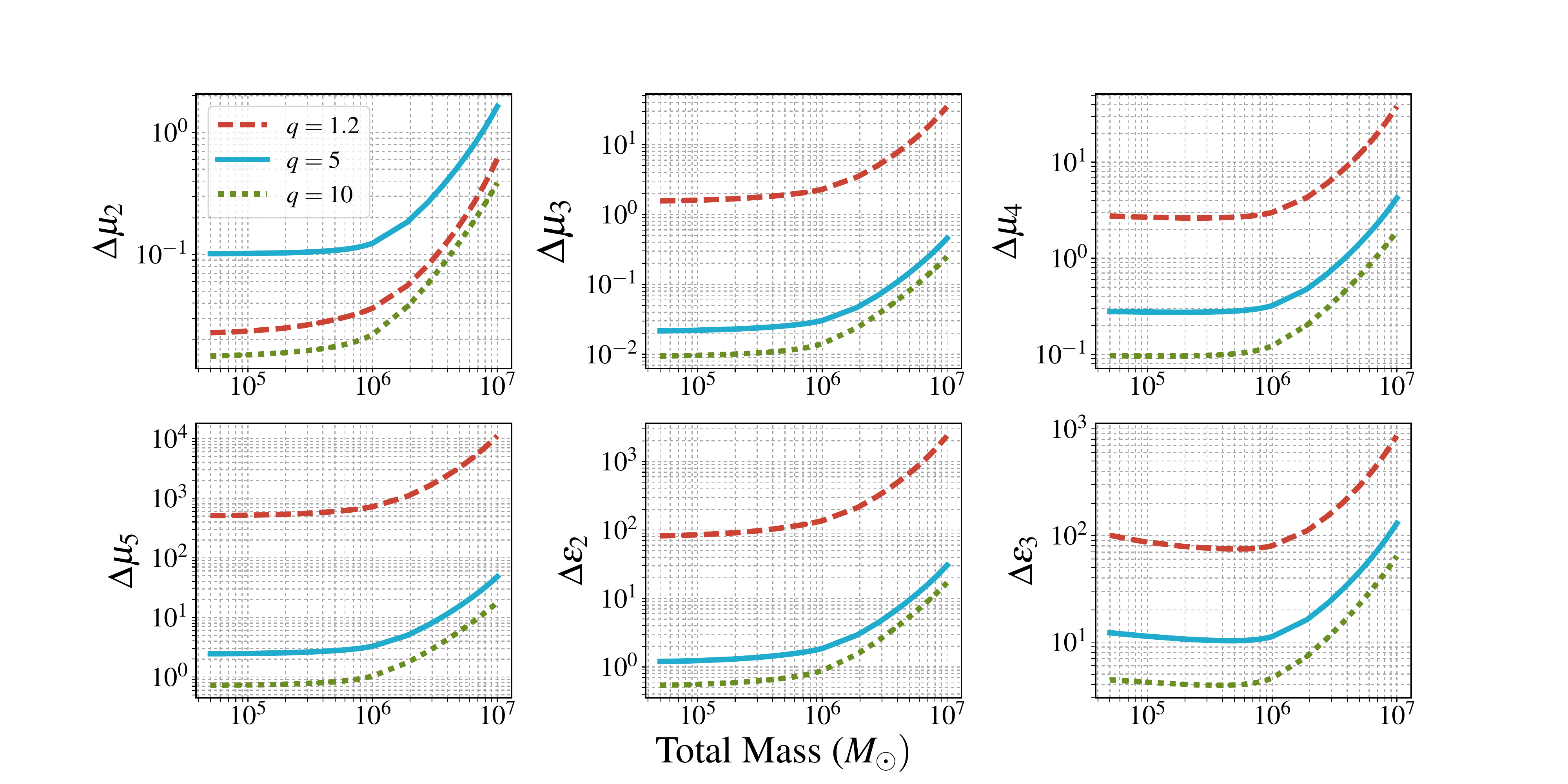}
	\caption{Projected $1\sigma$ errors on the multipole coefficients as a function of total mass for three different mass ratios $q=m_1/m_2=1.2,5$ and $10$ in case of LISA noise PSD. We assume $\chi_1=0.3, \chi_2=0.2$. All the sources are considered to be at a fixed luminosity distance of 3 Gpc. To obtain the numerical estimates showed in this plot, we also consider a prior distribution on $\phi_c$. To be precise, we assume $\phi_c$ to follow a gaussian distribution with a zero mean and a variance of $1/\pi^2.$}\label{fig:sys_mu_lowspin}
\end{figure*}

\begin{figure*}[t]
	\centering
	\includegraphics[scale=0.45]{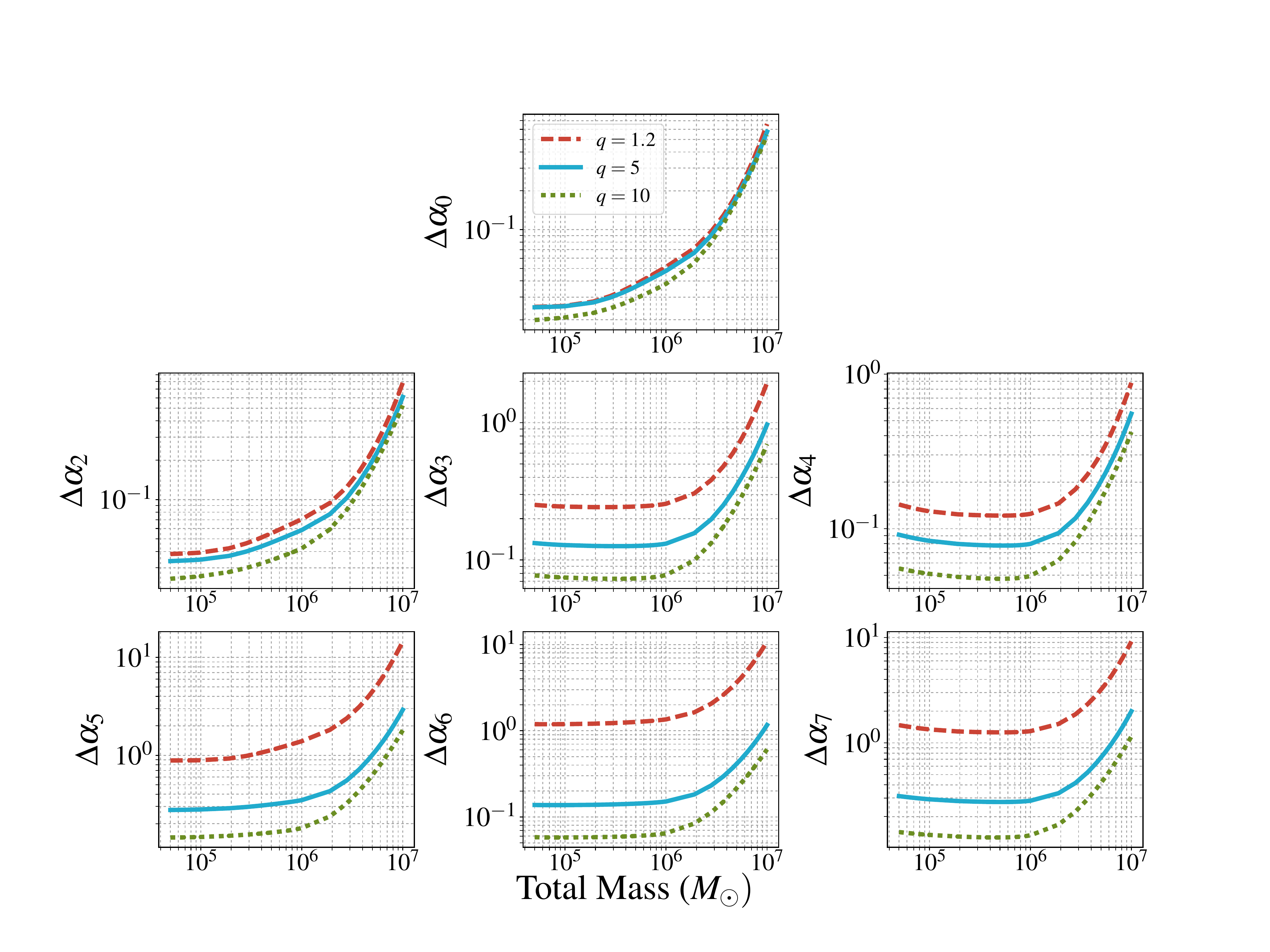}
	\caption{Projected $1\sigma$ errors on the energy coefficients as a function of total mass for three different mass ratios $q=m_1/m_2=1.2,5$ and $10$ in case of LISA noise PSD. We assume $\chi_1=0.9, \chi_2=0.8$. All the sources are considered to be at a fixed luminosity distance of 3 Gpc.  To obtain the numerical estimates showed in this plot, we also consider a prior distribution on $\phi_c$. To be precise, we assume the prior on $\phi_c$ to follow a gaussian distribution with a zero mean and a variance of $1/\pi^2.$}\label{fig:sys_alpha_highspin}
\end{figure*}

\begin{figure*}[t]
	\centering
	\includegraphics[scale=0.45]{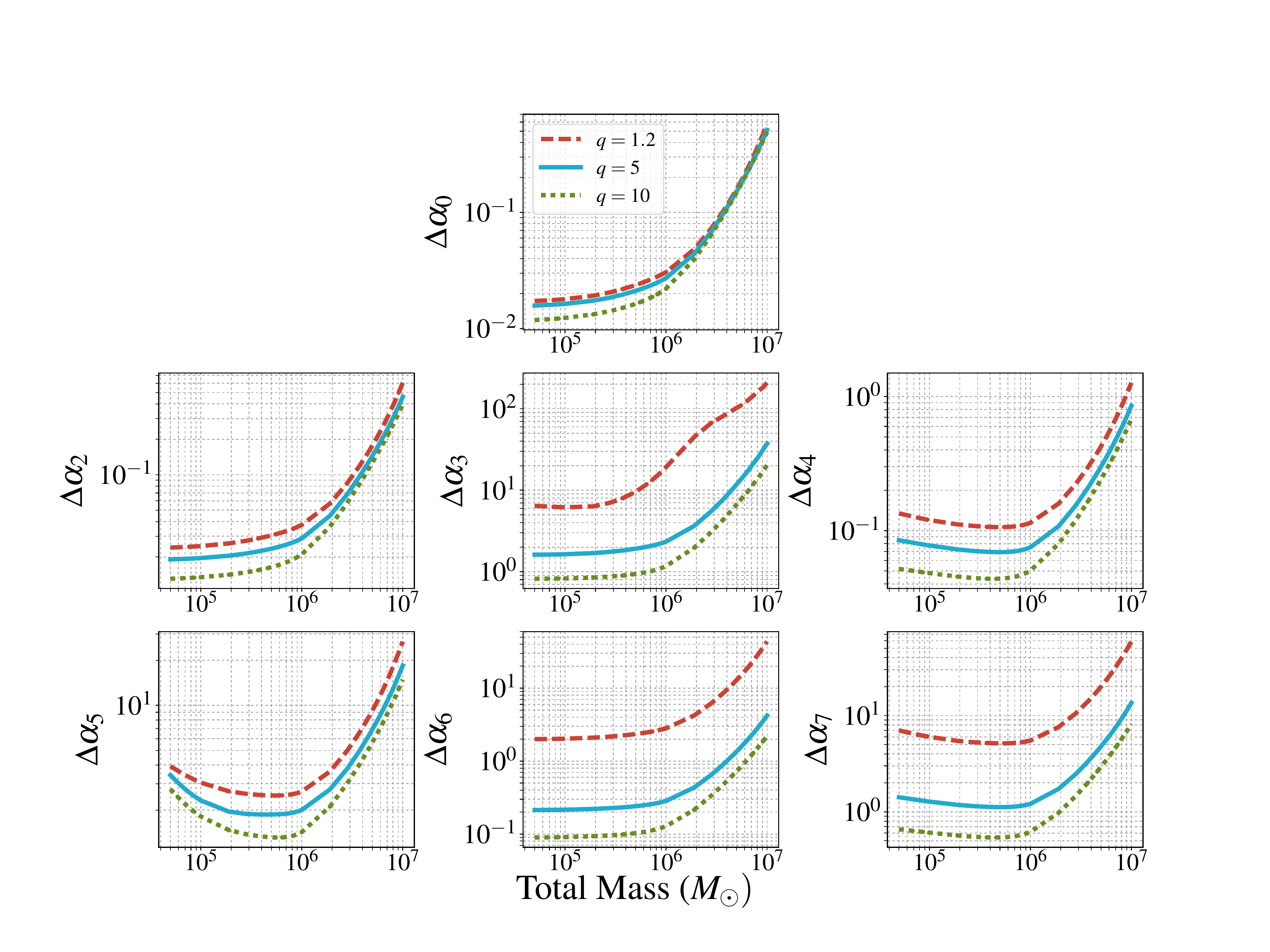}
	\caption{Projected $1\sigma$ errors on the energy coefficients as a function of total mass for three different mass ratios $q=m_1/m_2=1.2,5$ and $10$ in case of LISA noise PSD. We have considered $\chi_1=0.3, \chi_2=0.2$. All the sources are considered to be at a fixed luminosity distance of 3 Gpc. To obtain the numerical estimates showed in this plot, we also consider a prior distribution on $\phi_c$. To be precise, we assume the prior on $\phi_c$ to follow a gaussian distribution with a zero mean and a variance of $1/\pi^2.$}\label{fig:sys_alpha_lowspin}
\end{figure*}
\begin{figure*}[t]
	\centering
	\includegraphics[scale=0.45]{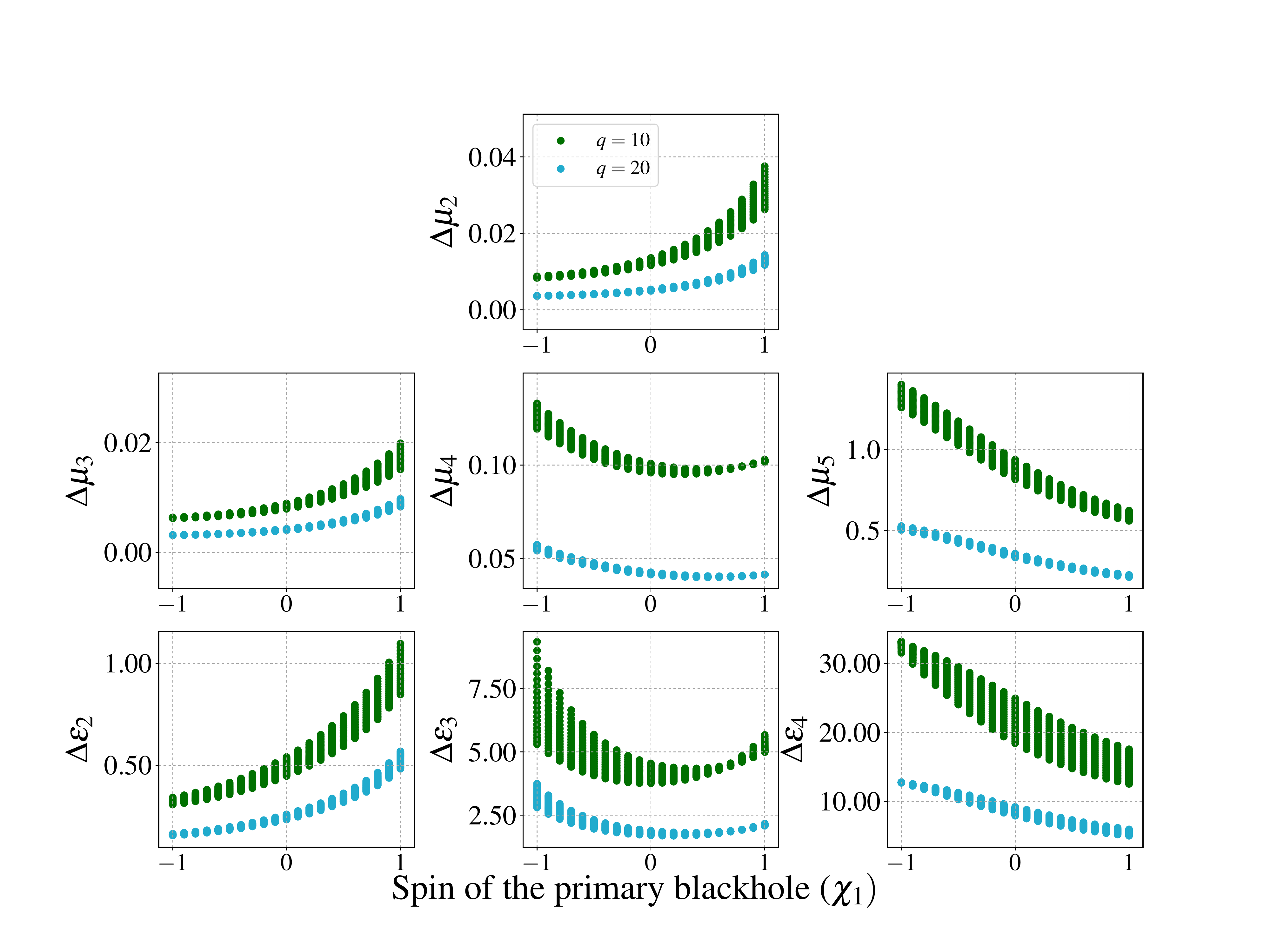}
	\caption{Projected $1\sigma$ errors on multipole coefficients as a function of the spin of the heavier black hole, $\chi_1$, for LISA noise PSD. All the sources are considered to be at a fixed luminosity distance of 3 Gpc with a total mass of $2\times 10^5$ \msun. The green dots are for mass ratio 10 and the cyan dots denotes mass ratio 20. The vertical spread in the bounds at each $\chi_1$ value is due to different $\chi_2$ in the range $[-1,1]$.To obtain the numerical estimates showed in this plot, we also consider a prior distribution on $\phi_c$. To be precise, we assume the prior on $\phi_c$ to follow a gaussian distribution with a zero mean and a variance of $1/\pi^2.$ }\label{fig:sys_spin}
\end{figure*}

\section{Detector Configurations}\label{detector config}
We describe here the various detector configurations we considered in
the present study.
\subsection{Ground-based second generation detector network}\label{2G-section}
As a representative case, we consider a world-wide network of five second-generation ground based detectors:
LIGO-Hanford, LIGO-Livingston, Virgo, KAGRA \cite{KAGRA_ref}, and LIGO-India \cite{Ligo-india}.  
We assume the noise PSD  for LIGO-Hanford, LIGO-Livingstone and LIGO-India to be the analytical fit 
given in Ref.~\cite{Ajith2011} whereas the following fit is used for Virgo PSD,

\begin{align}
  S^{\text{virgo}}_{h}(f)&=1.5344\times 10^{-47}\Bigg[1+1871\times\Bigg(\frac{16}{f}\Bigg)^{10} +11.72\times\Bigg(\frac{30}{f}\Bigg)^{6} 
  \nonumber \\ &
  + 0.7431\times\Bigg(\frac{50}{f}\Bigg)^{2}+ 0.9404\times\Bigg(\frac{70}{f}\Bigg) 
  + 0.2107\times\Bigg(\frac{100}{f}\Bigg)^{0.5}  \nonumber \\ &
  +26.02\Bigg(\frac{f}{500}\Bigg)^2
  \Bigg] \, {\rm Hz^{-1}}\,,
\end{align}
where $f$ is in units of Hz.
We consider the lower cut off frequency $\text{f}_{\text{low}}=10$ Hz for these detectors. 
For the Japanese detector KAGRA we use the noise PSD given in Ref.~\cite{SensitivityCurves} 
with $\text{f}_{\text{low}}=1$ Hz. For all the detectors, $f_{\rm high}$ is 
taken to be the frequency at the last stable orbit, $f_{\rm LSO}=1/(\pi\,m\,6^{3/2})$. 
As opposed to the single detector Fisher matrix analysis, for a network of detectors, 
Fisher matrix is evaluated for each detector and then added to obtain the network-Fisher-matrix. 
To estimate the individual Fisher matrices we use a waveform that is weighted with the correct antenna pattern 
functions $F_{+/\times}(\theta,\phi,\psi)$ of the detectors, 
where $\theta, \phi$ and $\psi$ are the declination,  the right ascension and the polarization angle of the source in the sky. 
More precisely we use the following waveform

\begin{align}\label{2G-waveform}
\tilde{h}(f)&=\frac{1+\cos^2 \iota}{2}\, F_+(\theta,\phi,\psi) \,\tilde{h}_+(f) \nonumber \\
            &+ \cos\iota \,F_\times(\theta,\phi,\psi)\, \tilde{h}_\times(f)
\end{align}
with
\begin{align}
\tilde{h}_+(f)&={\cal A}\, \mu_2 \, f^{-7/6}e^{-i\Psi_s}\,,\\
\tilde{h}_\times(f)&=-i\,\tilde{h}_+(f)\,.
\end{align}
The individual $F_{+/\times}(\theta,\phi,\psi)$ for each detector are estimated incorporating their location on Earth and Earth's rotation as given in Ref.~\cite{LAL_antenna_pattern}. 
We calculate the Fisher matrix for each detector considering an eight dimensional parameter space, $\{ t_c, \phi_c, \log \mathcal{A}, \log\mathcal{M}_c, \log\nu, \chi_s, \chi_a, \mu_\ell$ or $\epsilon_\ell$ or $\alpha_m\}$ specifying the GW signal. 
Here we fix the four angles, $\theta,\phi,\psi,\iota$ to be
$\pi/6,\pi/3,\pi/6,\pi/5$ respectively and do not treat them as
parameters in the Fisher matrix estimation. 
These four angles, being the extrinsic parameters, have small correlation with the intrinsic ones especially with the multipole or the energy coefficients, and hence have negligible effect on their measurement.

\subsection{Ground-based third generation detector network}
As a representative case for the third generation ground-based detector network, we consider three detectors: one Cosmic Explorer-wide band (CE-wb)~\cite{CE_WB} in Australia, one CE-wb in Utah-USA and one Einstein Telescope-D (ET-D)~\cite{ET-D} in Europe. We use the noise PSD given in Ref.~\cite{ET-D} for ET-D and the analytical fit given in Ref.~\cite{Kastha2018} for the CE-wb. 
%
%
We assume $f_{\rm low}$ to be $1$ and $5$ Hz for the ET-D and CE-wb,
respectively. To evaluate the Fisher matrix for this network
configuration we use the same waveform as given in
Eq.~(\ref{2G-waveform}) except for the estimation of Fisher matrix in
case of ET-D, we multiply the waveform by $\sin(\pi/3)$ because of its
triangular shape. We follow the same scheme as described in
Sec.~\ref{2G-section} to estimate the $1\sigma$ bounds on
$\mu_2,\mu_3,\mu_4,\epsilon_2$ and $\alpha_0,\alpha_2,\alpha_3,
\alpha_4$.

\subsection{Space-based LISA detector}

For the space based detector, LISA, we use analytical fit given in \cite{Babak2017} and choose $f_{\rm low}$ in such a
way that the signal stays in the detector band for  one year or less
depending on the frequency at the last stable orbit. More specifically,
we assume $f_{\rm low}$ to be~\cite{Cutler98,BBW05a} 

\begin{equation}
f_{\rm low}= \rm {max}\Bigg[10^{-5}, 4.149\times 10^{-5} \Bigg(\frac{\mathcal{M}_c}{10^6 M_\odot}\Bigg)^{-5/8}\Bigg(\frac{T_{\rm obs}}{1 \rm yr}\Bigg)^{-3/8}\Bigg],
\end{equation}
where $T_{\rm obs}$ is the observation time  which we consider to be
one year. We assume the upper cut off frequency, $f_{\rm high}$, to be the minimum of $[0.1, f_{\rm LSO}]$.  The waveform we employ for LISA is given in Eq.~(\ref{mod-TaylorF2}) except we multiply it by an additional factor of $\sqrt{3}/2$ in order to account for the triangular shape of the detector. We do not account for the orbital motion of LISA in
our calculations and consider LISA to be a single detector. 

We next discuss the Fisher matrix projections for the various
deformation coefficients parametrizing the conservative and dissipative
sectors in the context of advanced ground-based and space-based
gravitational wave detectors.
\section{Results}\label{sec:results}

 Our results for the ground-based detectors are depicted in Figs.
\ref{fig:2G_network_bounds} (second generation) and
~\ref{fig:3G_network_bounds} (third generation) and those for the
space-based LISA detector are presented in Figs. \ref{fig:sys_mu_highspin},
\ref{fig:sys_mu_lowspin}, \ref{fig:sys_alpha_highspin},
\ref{fig:sys_alpha_lowspin} and \ref{fig:sys_spin}. 
For the second and third generation ground-based detectors
configurations, we choose the binary systems with two different mass ratios $q=1.2, 5$
for two sets of spin configurations: high spin case
($\chi_1=0.9,\chi_2=0.8$) and low spin case ($\chi_1=0.3,\chi_2=0.2$).
We also assume the luminosity distance to all these prototypical sources to be 100 Mpc.
 We
consider these sources are detected with a network of second or third
generation detectors as detailed in the last section. For LISA, we
consider our prototypical supermassive BHs  to be at the luminosity distance of 3 Gpc with three different
mass ratios of $q=1.2, 5, 10$. For these mass ratios, we investigate both
high spin ($\chi_1=0.9, \chi_2=0.8$) and low spin ($\chi_1=0.3,
\chi_2=0.2$) scenarios. 

First we discuss the
qualitative features in the plots. As expected, the third generation
detector network which has better band width and sensitivity does better than
the second generation detectors whereas LISA and third generation
detectors perform comparably, though for totally different source
configurations.  The bounds on the
multipole coefficients describing the dissipative dynamics broadly
follows the trends seen in the non-spinning study carried out in
Ref.~\cite{Kastha2018}.  The mass-type multipole moments are 
measured better than the current-type ones appearing at the same PN order with $\mu_2$ (corresponding to the mass
quadrupole) yielding the best constraint as it is the dominant multipole
which contribute to the flux and the phasing. Due to the interplay
between the sensitivity and mass dependent upper cut-off frequency, the
errors increase as a function of mass in the regions of the parameter space we
explore. The errors improve as the mass ratio increases for all cases
except $\mu_2$. As argued in Ref.~\cite{Kastha2018}, $\mu_2$ is the only
multipole parameter which appears both in the amplitude and the phase of
the waveform and hence shows trends different from the other multipole
coefficients. Inclusion of spins, on
the whole, worsens the estimation of the multipole coefficients compared
to the non-spinning case. This is expected as the spins increase the
dimensionality of the parameter space but does not give rise to new
features that helps the estimation. Effects such as spin-induced
precession, which bring in a new time scale and associated modulations,
may help counter this degradation in the parameter estimation. But this will be a topic for a future
investigation. We also find that  as a function of the spin magnitudes, the parameter estimation
improves and hence highly spinning systems would yield stronger
constraints on these coefficients. The estimation of various $\alpha_k$,
parametrizing the conservative dynamics, also broadly follow these
trends. However, there is an important exception. The bounds on
$\alpha_3$ is consistently worse than those of $\alpha_4$. This may be
attributed to the important difference between them that $\alpha_3$
parametrizes the 1.5PN term in the conserved energy which has only
spin-dependent terms  whereas the 2PN term contains both non-spinning and
spinning contributions. Hence though $\alpha_4$ is sub-leading in the
PN counting,
and hence the bounds are better.

We now discuss the quantitative results from these plots. One of the
most interesting results is the projected constraints on coefficients
that parametrize conservative
dynamics. For third generation ground-based detectors, and for the
prototypical source specifications, the bounds on 2PN conservative
dynamics can be $\sim 10^{-2}$ which is comparable to or even better than the corresponding
bounds expected from LISA. On the multipole coefficients side, the quadrupole
coefficient $\mu_2$ may be constrained to $\leq 10^{-1}$($10^{-2}$) for
second (third) generation detector network while the bounds from LISA
are also $\sim 10^{-2}$. The best bounds for $\mu_3$ are $\sim 10^{-1},
10^{-2}, 10^{-2}$ for second generation, third generation and LISA,
respectively, corresponding to highly spinning binaries. The
projected bounds on the higher multipole coefficients from third
generation detector network and LISA are comparable in all these cases, though
one should keep in mind the specifications of the sources we consider
for these two cases are very different.

\section{Conclusion}\label{sec:conc}

We extend our previous work ~\cite{Kastha2018} by including spin effects
in the inspiral dynamics and provide a waveform model, parametrized in
terms of multipole and PN binding energy coefficients, for non-precessing compact binaries in
quasi-circular orbit. The spin-orbit contributions are computed up to
3.5PN order while the spin-spin contributions are obtained up to 3PN
order.
We also provide the projected $1\sigma$ bounds on the multipole coefficients as well as the PN deviation parameters in the conserved energy for the second generation ground based detector network, the third generation ground based detector network and the 
space-based detector LISA, using the Fisher matrix approach. We find that the four leading order multipole coefficients and the four leading order PN conserved energy coefficients are measured with reasonable accuracy using these GW detectors.

We are currently in the process of implementing this parametrized waveform model presented in this paper in  {\it LALInference} \cite{lalinference} to carry out tests of GR proposed here on real GW data.
As a follow up, it will be interesting to compute the parametrized
waveform within the effective-one-body formalism and investigate the
possible bounds on these coefficients. Inclusion of higher modes of
the gravitational waveforms, which contain these multipole coefficients
in the amplitude of the waveform, will also be an interesting follow up
in the future.

\section{Acknowledgment}
SK and KGA thank B. Iyer, G. Date, A. Ghosh and J. Hoque for
several useful discussions and  N. V. Krishnendu for cross-checking
some of the calculations reported here.
We thank B. Iyer for critical reading of the manuscript and providing useful comments.
KGA, AG, SK and BSS acknowledge the support
by the Indo-US Science and Technology Forum through the Indo-US {\em
	Centre for the Exploration of Extreme Gravity}, grant
IUSSTF/JC-029/2016. AG and BSS are supported in part by NSF grants
PHY-1836779, AST-1716394 and AST-1708146.  KGA is partially support by a
grant from Infosys Foundation. KGA also acknowledge partial support from
the grant EMR/2016/005594 by SERB. CVdB is supported by the research programme of the Netherlands Organisation for Scientific Research (NWO). Computing resources for this project were provided by the Pennsylvania State University.  This document has LIGO preprint number {\tt LIGO-P1900136}. 
 
\appendix*
\section{Systematic bias due to the use of non-spinning waveform model for GW detections by planned space-based detector LISA} 

\begin{figure*}[t]
	\centering
	\includegraphics[scale=0.38]{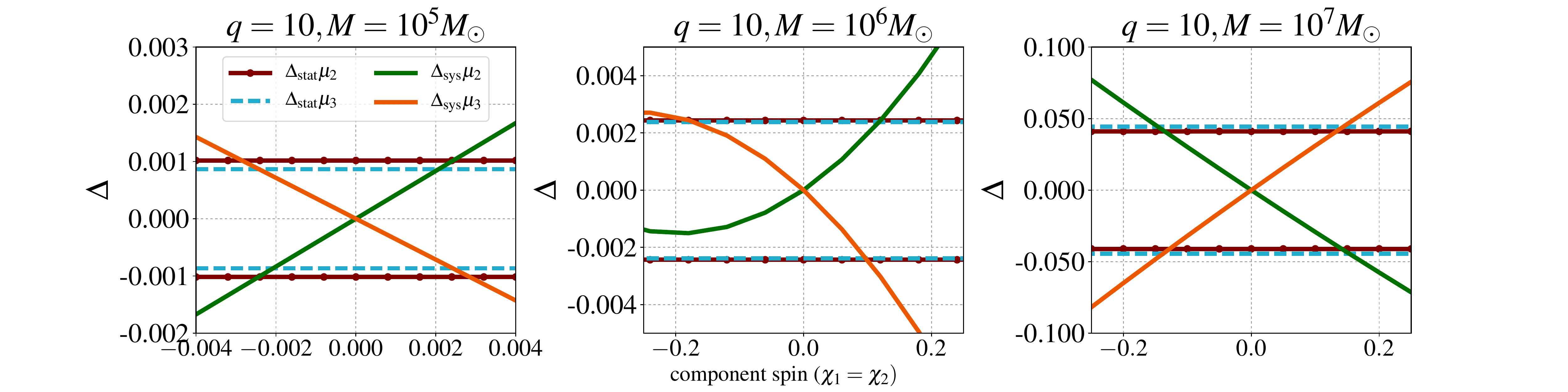}
	\caption{Numerical estimates of systematic biases on the  two leading multipole coefficients $\mu_2$ and $\mu_3$ as a function of $\chi_1=\chi_2=\chi$ for LISA noise PSD. We consider systems with three different total masses, $m=10^5, 10^6, 10^7 M_\odot$ having mass ratio $q=10$. All the sources are considered to be at a fixed luminosity distance of 3 Gpc.}
	\label{fig:bias}
\end{figure*}

The use of inaccurate
waveform model may lead to systematic biases in the parameter
estimation~\cite{CutlerVallisneri07,Favata2013}. For a detector data stream, s, consisting of a true waveform $\tilde{h}_{\rm{T}}(f;\vec{\theta}^{\rm{T}})$ and recovered with an approximate waveform $\tilde{h}_\text{AP}(f;\vec{\theta}^{\text {best fit}})$, the systematic errors on various parameters can be obtained by minimizing 
$\Big\langle [\tilde{h}_{\rm{T}}(f;\vec{\theta}^{\rm{T}})-\tilde{h}_\text{AP}(f;\vec{\theta}^{\text {best fit}})],[\tilde{h}_{\rm{T}}(f;\vec{\theta}^{\rm{T}})-\tilde{h}_\text{AP}(f;\vec{\theta}^{\text {best fit}})] \Big\rangle$~\cite{CutlerVallisneri07}.
Since we are interested in quantifying the systematics due to the
difference between the spinning and non-spinning waveforms, we adopt the
minimization scheme developed in Ref.~\cite{CutlerVallisneri07}. The
basic assumption behind this scheme is to define a one parameter family
of waveform models $(\tilde{h}^\lambda(f;\theta))$ that interpolate
between both $\tilde{h}_{\rm{T}}(f;\vec{\theta}^{\rm{T}})\equiv
\tilde{h}^{\lambda=1}(f;\theta)$ and
$\tilde{h}_\text{AP}(f;\vec{\theta})\equiv
\tilde{h}^{\lambda=0}(f;\theta)$. As it turns out, after a set of approximations, the linearized estimate for the systematic error is (see Eq.~(29) in Ref.~\cite{CutlerVallisneri07})

\begin{equation}
\Delta_{\text{sys}}\theta_m=\Big(\Gamma_{\rm AP}^{-1}\Big)_{mk}\Big\langle i\mathcal{A}\mu_2 f^{-7/6}\Delta\psi e^{i\psi}\Big|_{\theta=\theta^{\rm best \,fit}} ,\frac{\partial\tilde{h}_\text{AP}(f;\vec{\theta}^{\text {best fit}})}{\partial\theta_k}\Big\rangle \,,
\end{equation} 
where $(\Gamma_{\text AP})_{mk}$ is the Fisher matrix obtained from the approximate waveform $\tilde{h}_\text{AP}(f;\vec{\theta})$ and  $\Delta\psi=\psi_T-\psi_{\rm AP}$. All the quantities are evaluated at the best fit values of the parameters which coincide with the true values in the large SNR limit.

To quantify the systematic bias, we consider
a six dimensional parameter space consists of $\{t_c, \phi_c, {\rm
ln}\mathcal{A}, {\rm ln}\mathcal{M}_c, {\rm ln}\nu, \mu_\ell$ or
$\epsilon_\ell\}$ to completely specify the approximate waveform
$\tilde{h}_\text{AP} (f;\vec{\theta}^{\text {best fit}})$, for our
purpose the parametrized non-spinning TaylorF2 waveform. We use the same approximate
waveform to estimate the six dimensional Fisher matrix, $\Gamma_{\rm AP}.$
On the other hand, we consider the parametrized non-precessing TaylorF2 waveform to
be our true waveform model.

 In Fig.~\ref{fig:bias} we show the systematic biases on  $\mu_2$ and
$\mu_3$ for binaries with three different total masses, $M=10^5$ \msun, $10^6$ \msun,
$10^7$ \msun~and mass ratio $q=10$ as a function of individual spin
parameter $\chi_1=\chi_2=\chi$ for LISA. Due to a smaller total mass ($M=10^5$\msun) a large number of inspiral cycles reside in the LISA band. Hence even with very small spin values $\chi\sim\mathcal{O}(10^{-3})$, the systematic errors become larger than the statistical errors, which demands a parametrized spinning waveform model.
In contrast, for larger total masses of about $10^6$ \msun~or $10^7$ \msun, the systematics affect the parameter estimation when the spin magnitude is slightly larger $\sim\mathcal{O}(10^{-1})$, as expected. Hence it is very crucial to incorporate the spin corrections in the waveform to reduce the effects of systematics when extracting the information about the multipole coefficients. We also find that as the total mass of binary increases the slope of the systematic bias curves changes from positive to negative for $\mu_2$ and vice-versa for $\mu_3$. This could be due to the nature of the correlation  (positive or negative) between these multipole coefficients and the binary parameters (such as masses and spins) with increasing total mass. 
We quote the leading order estimates for the systematic biases in case of LISA only. Since the Fisher matrix-based leading order estimation of systematic biases for network configuration demands reformulation of the prescription, we postpone these for future study in a more rigorous and accurate Bayesian framework.

We give the inputs needed to compute the phasing for TaylorT2, TaylorT3 and
TaylorT4 in a Mathematica file (supl-Multipole-spin.m) which
serves the Supplemental Material to this paper.

\bibliographystyle{apsrev}
\bibliography{./Multipole_spin.bbl}

\end{document}